\documentclass[reprint,amsmath,amssymb,aps]{revtex4-1}

\usepackage{graphicx}
\usepackage{natbib}
\usepackage{amsmath}
\usepackage{hyperref}
\usepackage{gensymb}

\usepackage[usenames, dvipsnames]{color}

\newcommand\mnras{\rmfamily{Mon. Not. R. Astron. Soc.}}
\newcommand\aap{\rmfamily{Astron. Astrophys.}}
\newcommand\apjl{\rmfamily{Astrophys. J. Lett.}}

\newcommand\Beq{\begin{eqnarray}} 
\newcommand\Eeq{\end{eqnarray}}

\newcommand\Bfig{\begin{figure}} 
\newcommand\Efig{\end{figure}}

\newcommand\Pran{\mathrm{Pr}}

\renewcommand{\vec}[1]{\boldsymbol{#1}}

\newcommand{\grad}{\vec{\nabla}}
\renewcommand{\dot}{\vec{\cdot}}

\begin{document}

\title{Numerical Simulations of Internal Wave Generation by Convection in Water}

\author{Daniel Lecoanet}
\affiliation{%
Department of Astrophysics and Theoretical Astrophysics Center, University of California, Berkeley, CA 94720, USA\\
Kavli Institute for Theoretical Physics, University of California, Santa Barbara, CA 93106, USA}
\author{Michael Le Bars}
\affiliation{%
CNRS, Aix-Marseille Universit\'e, Ecole Centrale Marseille, IRPHE, Marseille, 13013, France\\
Department of Earth, Planetary, and Space Sciences,
University of California, Los Angeles, California 90095,
USA}
\author{Keaton J. Burns}
\affiliation{%
Department of Physics, Massachusetts Institute of Technology, Cambridge, MA 02139 USA\\
Kavli Institute for Theoretical Physics, University of California, Santa Barbara, CA 93106, USA}
\author{Geoffrey M. Vasil}
\affiliation{%
School of Mathematics \& Statistics, University of Sydney NSW 2006, Australia}
\author{Benjamin P. Brown}
\affiliation{%
Laboratory for Atmospheric and Space Physics and Department of Astrophysical \& Planetary Sciences, University of Colorado, Boulder, CO 80309\\
Kavli Institute for Theoretical Physics, University of California, Santa Barbara, CA 93106, USA}
\author{Eliot Quataert}
\affiliation{%
Department of Astrophysics and Theoretical Astrophysics Center, University of California, Berkeley, CA 94720, USA}
\author{Jeffrey S. Oishi}
\affiliation{%
Department of Physics, Farmingdale State College, Farmingdale, NY 11735\\
Department of Astrophysics, American Museum of Natural History, New York, NY 10024\\
Kavli Institute for Theoretical Physics, University of California, Santa Barbara, CA 93106, USA}

\date{\today}

\begin{abstract}
Water's density maximum at $4\degree {\rm C}$ makes it well suited to study internal gravity wave excitation by convection: an increasing temperature profile is unstable to convection below $4\degree{\rm C}$, but stably stratified above $4\degree{\rm C}$.  We present numerical simulations of a water-like fluid near its density maximum in a two dimensional domain.  We successfully model the damping of waves in the simulations using linear theory, provided we do not take the weak damping limit typically used in the literature.  In order to isolate the physical mechanism exciting internal waves, we use the novel spectral code Dedalus to run several simplified model simulations of our more detailed simulation.  We use data from the full simulation as source terms in two simplified models of internal wave excitation by convection: bulk excitation by convective Reynolds stresses, and interface forcing via the mechanical oscillator effect.  We find excellent agreement between the waves generated in the full simulation and the simplified simulation implementing the bulk excitation mechanism.  The interface forcing simulations over excite high frequency waves because they assume the excitation is by the ``impulsive'' penetration of plumes, which spreads energy to high frequencies.  However, we find the real excitation is instead by the ``sweeping'' motion of plumes parallel to the interface.  Our results imply that the bulk excitation mechanism is a very accurate heuristic for internal wave generation by convection.
\end{abstract}

\maketitle

\section{Introduction}

Internal waves are an important non-local transport mechanism in the Earth's atmosphere and in astrophysical fluids.  They transport quantities like energy, momentum, and angular momentum from where they are excited to where they damp.  Internal wave transport has been invoked to explain such disparate phenomena as the quasi-biennial oscillation (QBO) in the Earth's atmosphere \citep{baldwin01}; shear-layer oscillations at the solar tachycline \citep{talon02} and in other stars \citep{tc05}; mass loss during the last year of a massive star's life \citep{qs12,sq14}; and synchronization of the rotation of the cores of red giant branch stars \citep{fuller14}.

There are many sources of internal waves, including tidal interactions \citep[e.g.,][]{gl09}, flow over topography, loss of balance (geostrophic adjustment), and convection \citep{fa03}.  Here we focus on excitation via convection, which is an important source of wave excitation in stars and in the Earth's tropics.  Several heuristics are used to explain how internal waves are excited by convection; these include the obstacle effect, the mechanical oscillator effect, and bulk forcing \citep[e.g.,][]{as10}.

The obstacle effect describes the interaction of a horizontal flow over up-flowing plumes.  On time scales short in comparison to the plume time scale, the horizontal flow can generate internal waves in the same way that wind generates internal waves over topography \citep{clark86}.  We restrict ourselves to geometries where there are no mean horizontal flows in the stably stratified region, so the obstacle effect is not important in our simulations.  The mechanical oscillator effect refers to excitation by oscillations of the interface between the convective and stably stratified regions \citep{fovell92,as10}.  This is similar to the excitation of sound waves via oscillations in the surface of a drum.  In the bulk excitation picture, Reynolds stresses and thermal stresses within the convective region excite internal waves \citep{gk90,clark86}.  This is equivalent to the \citet{lighthill01} wave excitation theory, and has been applied extensively in the study of sound generation by turbulence \citep[e.g.,][and references within]{lesshafft10}.  The main purpose of this paper is to evaluate the strengths and deficiencies of the interface forcing and bulk excitation mechanisms.

A convenient system for studying convective excitation of internal waves is water near $4\degree{\rm C}$.  Since water has a density maximum at $4\degree{\rm C}$, a tank cooled to $0\degree{\rm C}$ at the bottom and kept near room temperature at the top is convectively unstable between $0\degree{\rm C}$ and $4\degree{\rm C}$ and stably stratified above $4\degree{\rm C}$.  This experiment was realized by \citet{townsend64, townsend66}, following a suggestion by Malkus.  The properties of the penetrative convection have been studied by various authors \citep[e.g.,][]{veronis63,brummell93}, but not as much attention has been paid to the wave excitation.

There are many numerical simulations of internal wave generation by convection.  \citet{mw73} simulate the water-ice system, and find evidence of internal gravity waves (IGWs).  More recently, several authors have studied wave generation in a stable layer adjoining a stratified convective region using the anelastic approximation \citep[e.g.,][]{rogers13,alvan14}.  Rather than simulate both stably stratified and convective regions, \citet{belkacem09} and \citet{sk13} have coupled simulations of convection (alone) to a theoretical model to estimate wave properties.

The rest of the paper is organized as follows.  In section~\ref{sec:full} we describe the numerical details of the main simulation.  Section~\ref{sec:characteristics} presents the broad characteristics of the convectively generated waves.  In section~\ref{sec:damping} we discuss the linear damping rates due to diffusion.  Sections~\ref{sec:source} \& \ref{sec:interface} describe simulations of the simulation.  We use data from the full simulation to test models of the bulk excitation and interface forcing heuristics.  We summarize the main results and conclude in section~\ref{sec:conclusion}.

\section{Full simulation}\label{sec:simexp}

\subsection{Numerical implementation}\label{sec:full}

We perform simulations of a fluid with a quadratic equation of state using the simple model of fresh water near $4\degree{\rm C}$ of \citet{veronis63}.  The simulations are run using Dedalus \citep[][see \url{dedalus-project.org} for more information]{burns15}, a general framework for studying partial differential equations, including eigenvalue problems, boundary value problems, and initial value problems (i.e., simulations). It uses a sparse Chebyshev spectral method to solve nearly arbitrary equation sets including algebraic constraints and complex boundary conditions. This flexibility allows us to simulate the water convection problem, as well as simplified linear models representing bulk excitation (section~\ref{sec:source}) and interface forcing (section~\ref{sec:interface}). In all cases, we use a two dimensional Cartesian domain with a Fourier grid in the horizontal ($x$) and a Chebyshev grid in the vertical ($z$) directions.  Many aspects of the simulation set-up are inspired by the quasi-two dimensional experimental investigation of \citet{lebars15}.

We solve the equations
\Beq
\grad\dot\vec{u} & = & 0, \label{continuity} \\
\partial_t \vec{u} + \grad p - \nu \nabla^2\vec{u} & = & - \vec{u}\dot\grad\vec{u} + \vec{g}\frac{\delta\rho}{\rho_0}, \label{momentum} \\
\partial_t T - \kappa \nabla^2 T & = & - \vec{u}\dot\grad T - k \left(T-T_{\rm air}\right), \label{temperature} \\
\frac{\delta\rho}{\rho_0} = \frac{\delta\rho(T)}{\rho_0} & = & -\alpha \left(T-T_0\right)^2, \label{equation of state}
\Eeq
where $\vec{u}, p, T$ are the velocity, pressure, and temperature, respectively.  $\delta\rho = \rho-\rho_0$ is the (small) density variation about a typical density of water, $\rho_0$.  $\nu, \kappa$ are the viscosity and thermal diffusivity, and $\vec{g}=-g\vec{e}_z$ is the gravitational acceleration.  Equation~\ref{continuity} is the continuity equation and equation~\ref{momentum} is the momentum equation.  Equation~\ref{temperature} is the temperature equation, and includes a Newtonian cooling term $k(T-T_{\rm air})$, which represents heat transfer to the ambient air in a third (not simulated) dimension.  The choice of $k$ sets the thermal equilibrium of the system: a small value of $k$ results in a large convective region with only a small stably stratified region, while a large value of $k$ results in a small convective region, with a very large stably stratified region.  Our choice of $k$ results in an equilibrium configuration with convective and stably stratified regions of approximately equal size, and corresponds to a heat transfer coefficient of 3.3 ${\rm W}/({\rm m}^2{\rm K})$.  The final equation~\ref{equation of state} is the equation of state, which is approximated to be quadratic around the density maximum at $T_0=4\degree{\rm C}$.  The diffusivities used in these studies are constant throughout the domain, and correspond to their values for water at $T=4\degree{\rm C}$; here $\Pran=\nu/\kappa=13.8$.  In contrast, the viscosity of water decreases by about a factor of two between $T=0$ and $25\degree{\rm C}$, while the thermal diffusivity stays about constant.

\begin{table}[b]
\caption{Parameter values used in the simulations. \label{tab:parameter values}}
\begin{ruledtabular}
\begin{tabular}{cc|cc}
parameter  & value & parameter & value \\[3pt] \hline
       $\nu$ & $1.8\times 10^{-6} \ {\rm m}^2/{\rm s}$ & $\kappa$ & $1.3\times 10^{-7} \ {\rm m}^2/{\rm s}$ \\
       $k$ & $2\times 10^{-5} \ {\rm s}^{-1}$ & $\alpha$ & $8.1\times 10^{-6} \ (\degree{\rm C})^{-2}$ \\
       $T_0$ & $4\degree{\rm C}$ & $T_{\rm air}$ & $21\degree{\rm C}$ \\
       $T_{\rm bot}$ & $0\degree{\rm C}$ & $T_{\rm top}$ & $25\degree{\rm C}$ \\
       $g$ & $9.8 \ {\rm m}/{\rm s}^2$ & $z_{\rm int}$ & $0.18 \ {\rm m}$\\
\end{tabular}
\end{ruledtabular}
\end{table}

The vertical extent of the domain is $0.35 \ {\rm m}$, and our vertical boundary conditions are no slip, and $T=T_{\rm bot}$, $T=T_{\rm top}$ on the bottom and top, respectively.  For the horizontally averaged mode, $\partial_z w=0$, where $w$ is the vertical velocity.  Thus, the conditions $w=0$ at the top and $w=0$ at the bottom are redundant.  We replace the latter boundary condition with the gauge choice $p=0$ at the bottom of the domain.  We use a resolution of 256 modes and grid points (with no dealiasing) in the vertical direction.  Dealiasing is not necessary because the solution is so well resolved that the explicit diffusivities are sufficient for reducing mode amplitudes at high wavenumber to very close to zero.  We have repeated portions of the calculation at double resolution, and found negligible differences in the statistics of the flow.

Our horizontal boundary conditions are $u=0$ and $\partial_xT=\partial_xw=\partial_x p=0$ on $x=0$ and $x=0.2 \ {\rm m}$.  This allows us to represent $u$ as a sine series in $x$, and $T$, $w$, and $p$ as cosine series in $x$.  Plumes rising on the sides of the domain do not exchange heat with the wall, so they lose their buoyancy more slowly than plumes in the center of the domain.  Thus, the most stable convective state has plumes rising on the sides of the domain.  We use a resolution of 512 sine/cosine modes and grid points (with no dealiasing) in the horizontal direction.  We do not dealias in the horizontal direction for the same reason that we do not dealias in the vertical direction.

The time integration uses a split implicit/explicit first-order scheme where certain terms are treated implicitly using forward Euler, and the remaining terms are treated explicitly using backward Euler.  We have repeated portions of the calculation with a third order, four-stage DIRK/ERK scheme \citep{ascher97}, and found negligible differences in the statistics of the flow.  The nonlinear terms, the buoyancy term, and the constant $k T_{\rm air}$ term in the temperature equation are treated explicitly, and all the other terms are evolved implicitly.  The time step is taken to be the lesser of either $0.1 \ {\rm s}$ or the advective Courant--Friedrichs--Lewy (CFL) time given by $0.08\times \min \left(\Delta x/u,\Delta z/w\right)$, where the minimum is taken over every grid point in the domain and $\Delta x$, $\Delta z$ are the grid spacing in the $x$ and $z$ directions, respectively.  Typical time steps are $\approx 0.02 \ {\rm s}$.

The initial temperature profile is piecewise linear, varying between $T_{\rm bot}$ at $z=0$ to $T_0$ at $z=z_{\rm int}=0.18 \ {\rm m}$, and then to $T_{\rm top}$ at $z=0.35 \ {\rm m}$ at the top of the domain.  We also add low amplitude random noise to the temperature field to initiate convection.  In this paper, we analyze the times between $35287 \ {\rm s}$ and $39122 \ {\rm s}$.  This corresponds to a period of about ten convective turnover times, starting about seventy convective turnover times after the beginning of the simulation.  We found that ten convective turnover times is long enough to build sufficient statistics to describe both the convection and waves.  Although we simulate many convective turnover times, the simulation up to $40000 \ {\rm s}$ corresponds to only 4\% of a thermal time, so the thermal structure continues to evolve on long time scales throughout the simulation.  Several dimensionless numbers describing the convection in this simulation are given in table~\ref{tab:dimensionless}.  Because our equations include a Newtonian cooling term, we do not have a constant heat flux through the convection zone.  Thus, we calculate the Nusselt number using the average value of $wT$ in the convection zone.

\begin{table}[b]
\caption{Dimensionless numbers characterizing the convection in the lower part of the domain.  $L_{\rm conv}$ is taken to be $0.22 \ {\rm m}$.  We use $\langle\cdot\rangle_{x, ...}$ to denote an average with respect to the variables listed in the subscript.  The $z$ average in the calculation of ${\rm Nu}$ is only within the convection zone.\label{tab:dimensionless}
}
\begin{ruledtabular}
\begin{tabular}{ccc|ccc|ccc}
&$\displaystyle{\rm Ra}=\frac{g\alpha T_0^2 L_{\rm conv}^3}{\nu \kappa}$ & & & $\displaystyle{\rm Re}=\frac{u_{\rm rms} L_{\rm conv}}{\nu}$ & & & $\displaystyle{\rm Nu}=\frac{-\langle wT\rangle_{x,z,t}}{\kappa T_0/L_{\rm conv}}$ & \\
\hline
&$\displaystyle5.8\times 10^{7}$ & & & 100 & & & 6 & \\
\end{tabular}
\end{ruledtabular}
\end{table}

\subsection{Characteristics of convectively generated waves}\label{sec:characteristics}

\begin{figure*}
  \centerline{\includegraphics{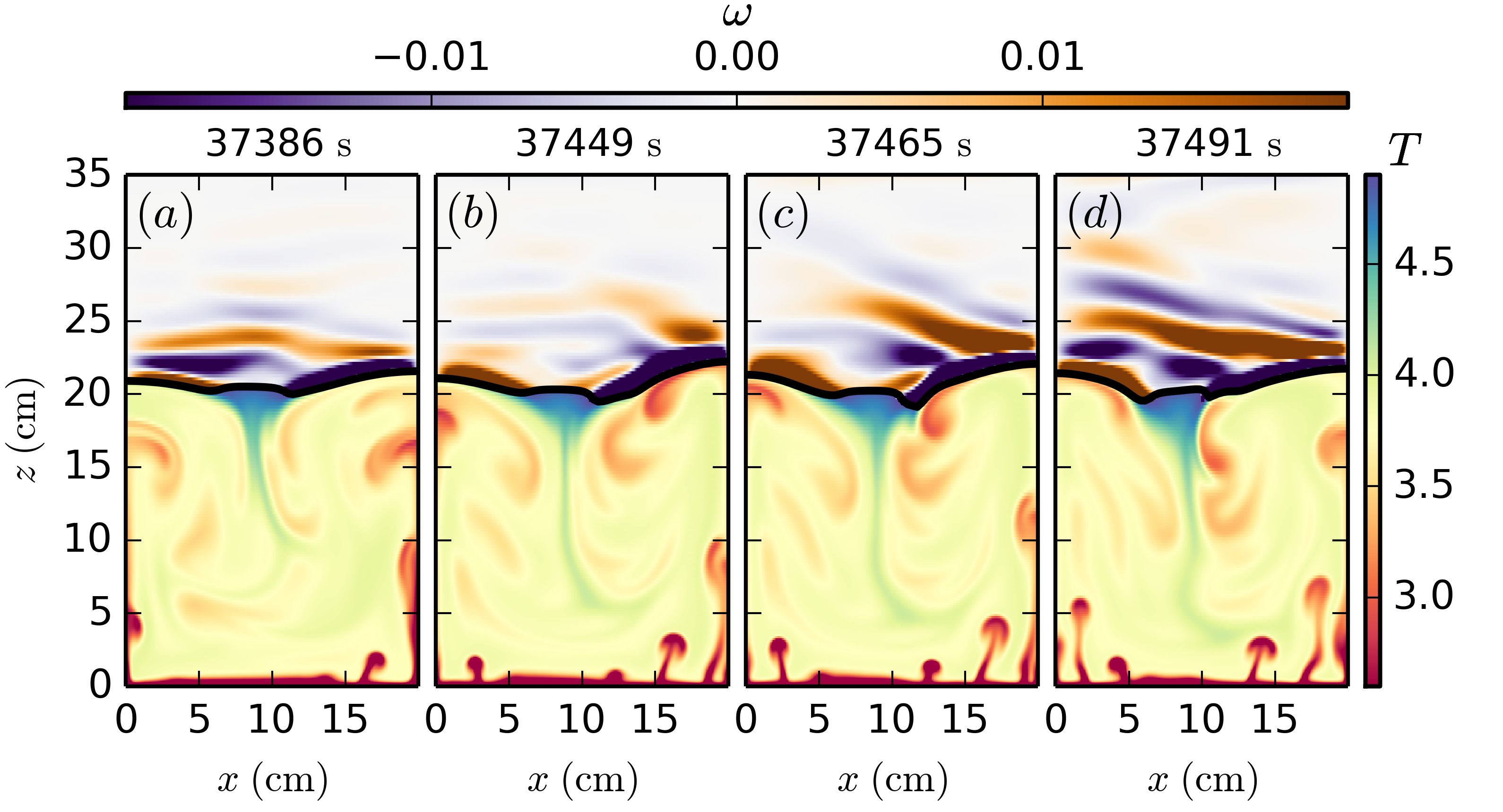}}
  \caption{(Color online) Four simulation snapshots near $t=37400 \ {\rm s}$.  The bottom part of the domain (below the thick black line) shows the temperature field.  Recall that below $4\degree{\rm C}$, cold water (red) is less dense than hot water (blue).  The thick black line is the $5\degree{\rm C}$ isotherm and shows the boundary between the convective region (below) and the stably stratified region (above).  The top part of the domain (above the thick black line) shows the vorticity field associated with IGWs.}
\label{fig:excitation}
\end{figure*}

\begin{figure*}
  \centerline{\includegraphics{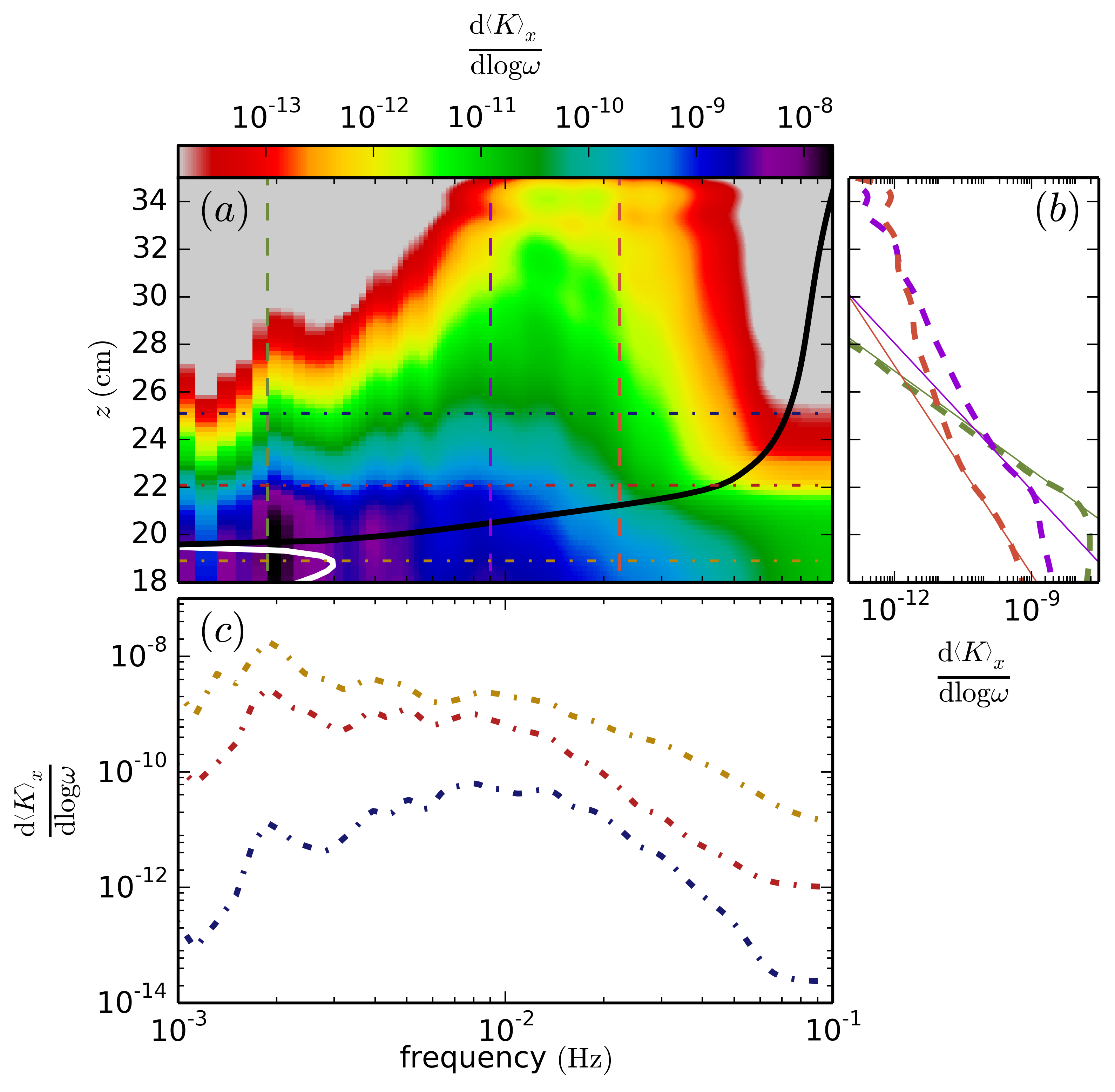}}
  \caption{(Color online) ($a$): spectrogram of the kinetic energy density.  The white curve is the imaginary part of the buoyancy frequency corresponding to the convection frequency.  The black curve is the real part of the buoyancy frequency in the stably stratified region.  The data has been smoothed to improve statistics (see text for more details).  Panels ($b$) \& ($c$) plot the kinetic energy density as a function of $z$ (dashed lines) \& $\omega$ (dot--dashed) respectively; the vertical and horizontal lines in panel ($a$) are color--coded to correspond to these slices.  The thin solid lines in panel ($b$) show the predicted linear damping rates near the convective--stably stratified boundary at $z=0.22 \ {\rm m}$.  Within the convection region (panel ($c$), yellow curve), the energy is peaked near the convection frequency.  However, these low frequency waves are quickly attenuated, and the stably stratified region is dominated with waves with frequencies $1$--$2\times 10^{-2} \ {\rm Hz}$ (panel ($c$), blue curve).}
\label{fig:overview}
\end{figure*}

The convective excitation of waves is depicted in figure~\ref{fig:excitation}.  Panel ($a$) shows the typical state before a major excitation event.  The convective region contains two counter-rotating convective cells, with up flows (represented by red cold, buoyant fluid) along the sides of the domain, and a down flow (represented by blue hot, dense fluid) in the center.  Because water has a $\Pran$ sufficiently higher than one, the thermal boundary layer at the bottom of the domain is unstable to the formation of buoyant plumes.  The turnover frequency of the convective cells ($\sim 2\times 10^{-3} \ {\rm Hz}$) and the plume ejection frequency ($\sim 10^{-2} \ {\rm Hz}$) are the important time scales in the convection zone.  The yellow curve in figure~\ref{fig:overview}, panel ($c$) shows the kinetic energy spectrum at the top of the convection zone ($z\approx 0.19 \ {\rm m}$).  Although there is a peak at the turnover frequency, the spectrum is fairly flat between the turnover frequency and the plume ejection frequency.  At frequencies higher than the plume ejection frequency, the spectrum falls off rapidly.

Panel ($a$) of figure~\ref{fig:excitation} shows a particularly vigorous plume on the right side of the domain at about $z=0.16 \ {\rm m}$.  This plume rises into the stably stratified region (water above $4\degree{\rm C}$) in panel ($b$), moving the interface between the convective and stably stratified regions upwards and generating strong, localized IGWs.  The plume is deflected by the stratified fluid above it, and deflects leftwards.  This allows the interface to lower (panels $c$ \& $d$).

Figure \ref{fig:excitation} shows the IGWs generated by the plume in the lower right corner of the stably stratified region propagate toward the upper-left.  However, the phase velocity is toward the lower-left---the upper part of the wave packet starts with positive vorticity (panel $b$), but shifts to having negative vorticity (panel $c$), and then positive vorticity again (panel $d$).  This is consistent with the IGW dispersion relation, which implies that the group velocity is perpendicular to the phase velocity.  The supplementary materials include a movie showing the flow evolution from $34630 \ {\rm s}$ to $39332 \ {\rm s}$.

IGWs are continually excited as the plumes detaching from the bottom boundary layer approach the interface.  Viscosity typically damps the waves before they can propagate to the top of the domain.  This is depicted in figure~\ref{fig:overview}($a$), a spectrogram of the kinetic energy density, i.e., a plot of $\omega \langle K\rangle_x \equiv 0.5 \ \omega \langle  |\vec{u}(\omega,x,z)|^2  \rangle_x$ as a function of $\omega$ and $z$, where $\langle\cdot\rangle_x$ denotes a horizontal average.  All logarithms in this paper are base $e$.  The frequency spacing of our data is uniform, so at high frequencies, the frequency density becomes large.  To improve the statistical power, we smooth the data by convolving with a gaussian with $\sigma_z=2.5\times 10^{-3} \ {\rm m}$ and $\sigma_{\log\omega}=0.05$.  Figure~\ref{fig:overview} ($a$) also shows the real (black line) and imaginary (white line) parts of the buoyancy (Brunt-V\"{a}is\"{a}l\"{a}) frequency, defined by
\Beq
N^2 = \alpha g \frac{{\rm d} \left(\left\langle T \right\rangle_{x,t}-T_0\right)^2}{{\rm d} z}.
\Eeq
The real part of the buoyancy frequency is the maximum frequency of IGWs, and the imaginary part gives the convection frequency within the convective region.

In the convective region (below $z\approx 0.22 \ {\rm m}$), the energy is peaked near the convection frequency at $f\approx 2\times 10^{-3} \ {\rm Hz}$ (figure~\ref{fig:overview} ($c$), yellow line).  The convection has power at a broad range of frequencies extending to high frequencies ($>10^{-1} \ {\rm Hz}$).  By contrast, the power in the stably stratified region is localized below the buoyancy frequency (figure~\ref{fig:overview} ($c$), red \& blue lines).  Although the excitation of internal waves is strongest near the convection frequency, these waves quickly viscously damp.  However, higher frequency waves with $f$ between $1$--$2\times 10^{-2} \ {\rm Hz}$ have longer damping lengths, and can propagate much further into the stably stratified region.

In a statistically steady state, the viscous damping of waves matches the convective excitation.  We use this to estimate the wave flux generated by convection and compare to theoretical predictions.  Equating the wave flux to the viscous damping, we find that the flux at a height $z$ in a mode with horizontal wavenumber $k_x$ and frequency $\omega$ is
\Beq\label{wave flux}
F_{\rm wave}(z, k_x,\omega) \sim \nu \ell_d(k_x,\omega) |\vec{k}|^2 |\vec{u}(z, k_x,\omega)|^2,
\Eeq
where $\ell_d$ is the linear viscous damping length.  We discuss the calculation of $\ell_d$ in section~\ref{sec:damping}.

To calculate the wave flux, we first calculate the kinetic energy of all modes with frequency $\omega$ as a function of height, $\left\langle|\vec{u}(z,\omega)|^2\right\rangle_{x}$.  Figure~\ref{fig:overview} ($b$) plots this for three different values of $\omega$.  Near the interface at $z=0.22 \ {\rm m}$, the kinetic energy is decreasing exponentially with height.  We find that this exponential decay is well fit by $\ell_d$ for a single $k_x$ typically corresponding to a large wavelength mode.  In calculating $\ell_d$, we assume that $N$ is constant and equal to $0.08 \ {\rm Hz}$, a typical value within the stably stratified region.  This suggests that most of the energy in waves with frequency $\omega$ is concentrated into a mode with a specific $k_x$ at $z=0.22 \ {\rm m}$.  In figure~\ref{fig:overview} ($b$), thin solid lines show the predicted damping for waves with $\omega=2\times 10^{-3} \ {\rm Hz}$ and wavelength $\lambda_x = 0.4 \ {\rm m}$ (green); $\omega=9\times 10^{-3} \ {\rm Hz}$ and $\lambda_x = 0.2 \ {\rm m}$ (purple); and, $\omega=2\times 10^{-2} \ {\rm Hz}$ and $\lambda_x = 0.1 \ {\rm m}$ (orange).  These all give good local fits, although the damping rate decreases with height for the two larger values of $\omega$.  This is because the waves with $\lambda_x = 0.1 \ {\rm m}$ or $0.2 \ {\rm m}$ damp more quickly than waves with $\lambda_x = 0.4 \ {\rm m}$, so although these lower wavelength waves are dominant at $z = 0.22 \ {\rm m}$, they become subdominant higher in the domain.

\begin{figure}
  \centerline{\includegraphics{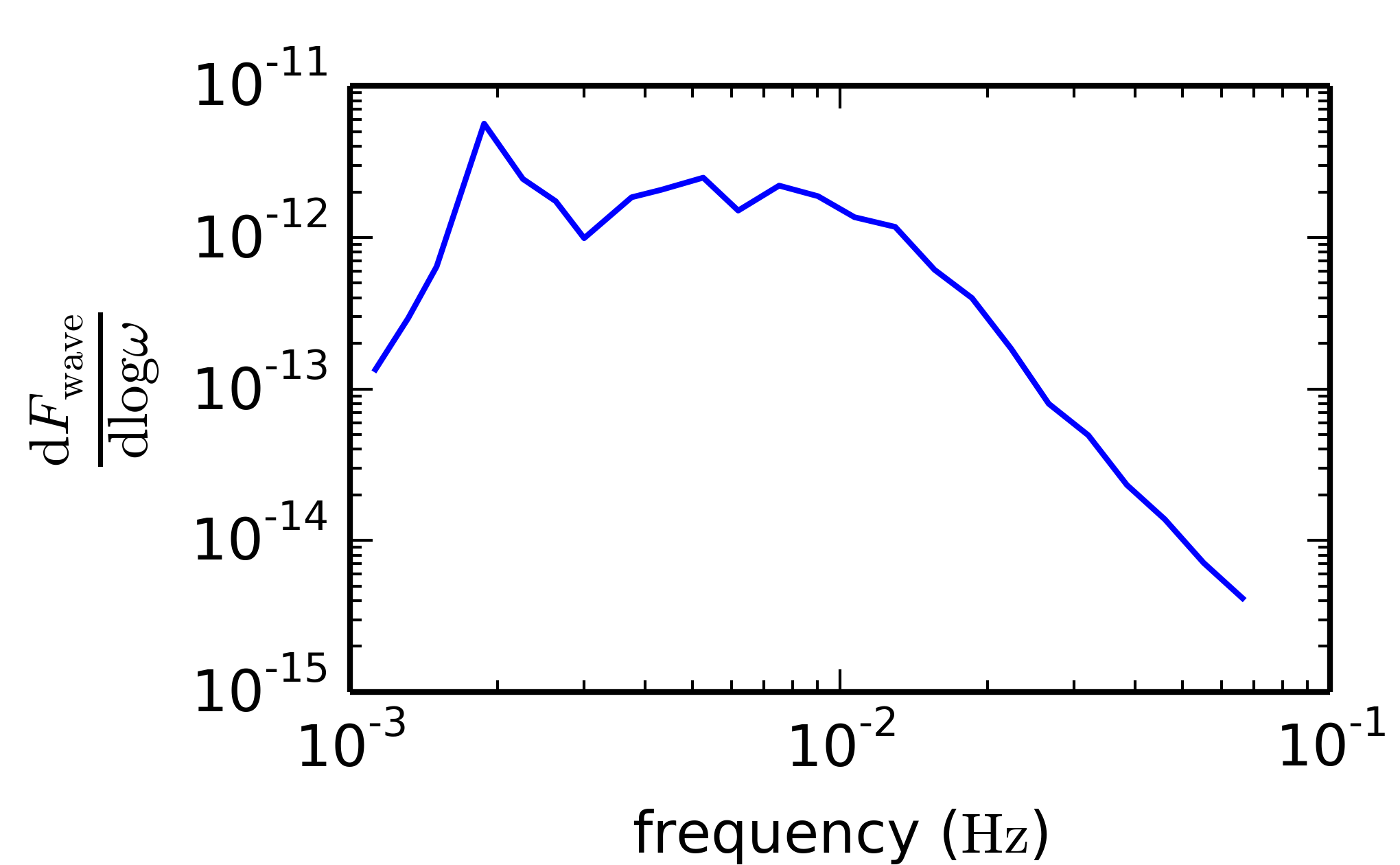}}
  \caption{(Color online) Wave flux spectrum at $z=0.22 \ {\rm m}$ calculated using equation~\ref{wave flux}.  We assume that the flux at each $\omega$ is dominated by a single $k_x$, which we determine by comparing the energy decay rate at $z=0.22 \ {\rm m}$ to $\ell_d(\omega,k_x)$.}
\label{fig:spectrum}
\end{figure}

This shows that the wave flux at $z = 0.22 \ {\rm m}$ (the position of the convective--stably stratified interface) for a given $\omega$ is dominated by a single $k_x$.  In figure~\ref{fig:spectrum}, we plot the flux as a function of $\omega$, where we assume all the energy is at this special value of $k_x$.  Although there is a peak in the wave flux spectrum at the turnover frequency ($2\times 10^{-3} \ {\rm Hz}$), the spectrum is fairly flat out to the plume ejection frequency ($10^{-2} \ {\rm Hz}$).  At higher frequencies, the wave flux decreases as $\omega^{-3}$.  The total flux is
\Beq
F_{\rm wave,sim} \sim 5\times 10^{-12} \ ({\rm m}/{\rm s})^3.
\Eeq

The theoretical prediction for the wave flux is $\mathcal{M} u_c^3$ \citep[e.g.,][]{gk90,lq13}, where $u_c$ is a typical convection velocity and $\mathcal{M}$, the ``convective Mach number,'' is an efficiency factor equal to the ratio of a typical convection frequency to a typical buoyancy frequency.  From figure~\ref{fig:overview} ($a$), we estimate $\mathcal{M}\sim 0.03$. If we estimate $u_c\sim \omega_c \ell_c$, where $\omega_c=2\times 10^{-3} \ {\rm Hz}$ is the turnover frequency and $\ell_c=0.22 \ {\rm m}$ is the depth of the convective region, then $u_c= 0.4\times 10^{-3} \ {\rm m}/{\rm s}$ and
\Beq
F_{\rm wave,theory,1} \sim 2.5\times 10^{-12} \ ({\rm m}/{\rm s})^3.
\Eeq
However, if we assume that $u_c$ is the rms velocity in the convection zone, this gives $u_c=0.9\times 10^{-3} \ {\rm m}/{\rm s}$ and
\Beq
F_{\rm wave,theory,2} \sim 2.5\times 10^{-11} \ ({\rm m}/{\rm s})^3.
\Eeq
Thus, the wave flux calculated in the simulations is in the range expected theoretically.  This calculation shows that the predicted flux depends sensitively on the assumptions made for various parameters, so we cannot expect more than rough agreement.

\section{Linear damping rates}
\label{sec:damping}

Wave damping is an important process in the simulation.  \citet{rm11,rogers13}, and \& \citet{alvan14} argued that the linear damping rate overestimates the damping in simulations of convectively excited waves.  They attribute the lack of damping to nonlinear interactions.  In contrast, we find that linear theory accurately describes the wave damping in our simulation, but only when the full damping rate is used.

The linear damping rate can be derived from the linearized equations of motion in the WentzelÐKramersÐBrillouin (WKB) limit.  \citet{ztm97} and \citet{press81} calculate the damping rate, but only in the quasi-adiabatic or weak dissipation limit, which is not always applicable.

We will describe the damping using an inverse damping length,
\Beq\label{damping factor}
E(z) = E(z_0) \exp\left(-2\int_{z_0}^z \ \ell_d^{-1} dz\right),
\Eeq
where we assume $z>z_0$.  In the weak dissipation limit, the inverse damping length due to viscosity is
\Beq\label{weak damping}
\ell_d^{-1} = \nu \frac{N^3 k_x^3}{\omega^4}.
\Eeq
For water, viscous damping dominates, while in astrophysical systems, $\Pran$ is typically much less than unity, so viscosity is typically less important than thermal diffusivity.  Equation~\ref{weak damping} shows that as $\omega$ goes to zero, the inverse damping length increases as $\omega^{-4}$.  This expression for the inverse damping length is valid only in the weak dissipation limit, i.e., when $\omega\gg \nu k^2$.  Geophysical and astrophysical systems have very small diffusivities, so this condition is satisfied.  However, laboratory experiments and numerical simulations have much larger diffusivities.  We find that most modes in our simulation are not in the weak dissipation limit, so we would not expect equation~\ref{weak damping} to represent the dynamics in the simulation.

The calculation of the linear damping rate in the WKB limit without assuming weak dissipation is more involved, but straightforward.  The inverse damping length is
\Beq\label{full damping}
\ell_d^{-1}&& =  \nonumber \\
&& {\rm Im}\left[ \frac{(-1)^{3/4}}{\sqrt{2}}\sqrt{-2ik_x^2 - \frac{\omega}{\nu} + \frac{\sqrt{\omega^3 + 4 i k_x^2\nu N^2}}{\nu \sqrt{\omega}}}\right]. \quad \quad
\Eeq
This reduces to equation~\ref{weak damping} in the weak damping limit, $\omega\gg k^2\nu$.  Furthermore, the vertical wavenumber also changes from the standard adiabatic result ($k_z=k_x N/\omega$),
\Beq
k_z && = \nonumber \\
&& {\rm Re}\left[ \frac{(-1)^{3/4}}{\sqrt{2}}\sqrt{-2ik_x^2 - \frac{\omega}{\nu} + \frac{\sqrt{\omega^3 + 4 i k_x^2\nu N^2}}{\nu \sqrt{\omega}}}\right]. \quad\quad
\Eeq
In the limit of small $\omega$, the inverse damping length in equation~\ref{full damping} approaches
\Beq
\ell_d^{-1} = \sin\left(7\pi/8\right) \left(\frac{k_x^2 N^2}{\nu\omega}\right)^{1/4}.
\Eeq
In this limit, the inverse damping length increases only as $\omega^{-1/4}$ as $\omega$ goes to zero, much more slowly than in the weak damping limit where $\ell_d^{-1}\sim \omega^{-4}$.  For small $\omega$, the weak damping limit predicts the waves are strongly damped.  This indicates that the weak damping limit is not appropriate.  We find that, although the actual wave damping is less severe than predicted by the weak damping limit, the waves are still strongly damped.

\begin{figure*}
  \centerline{\includegraphics{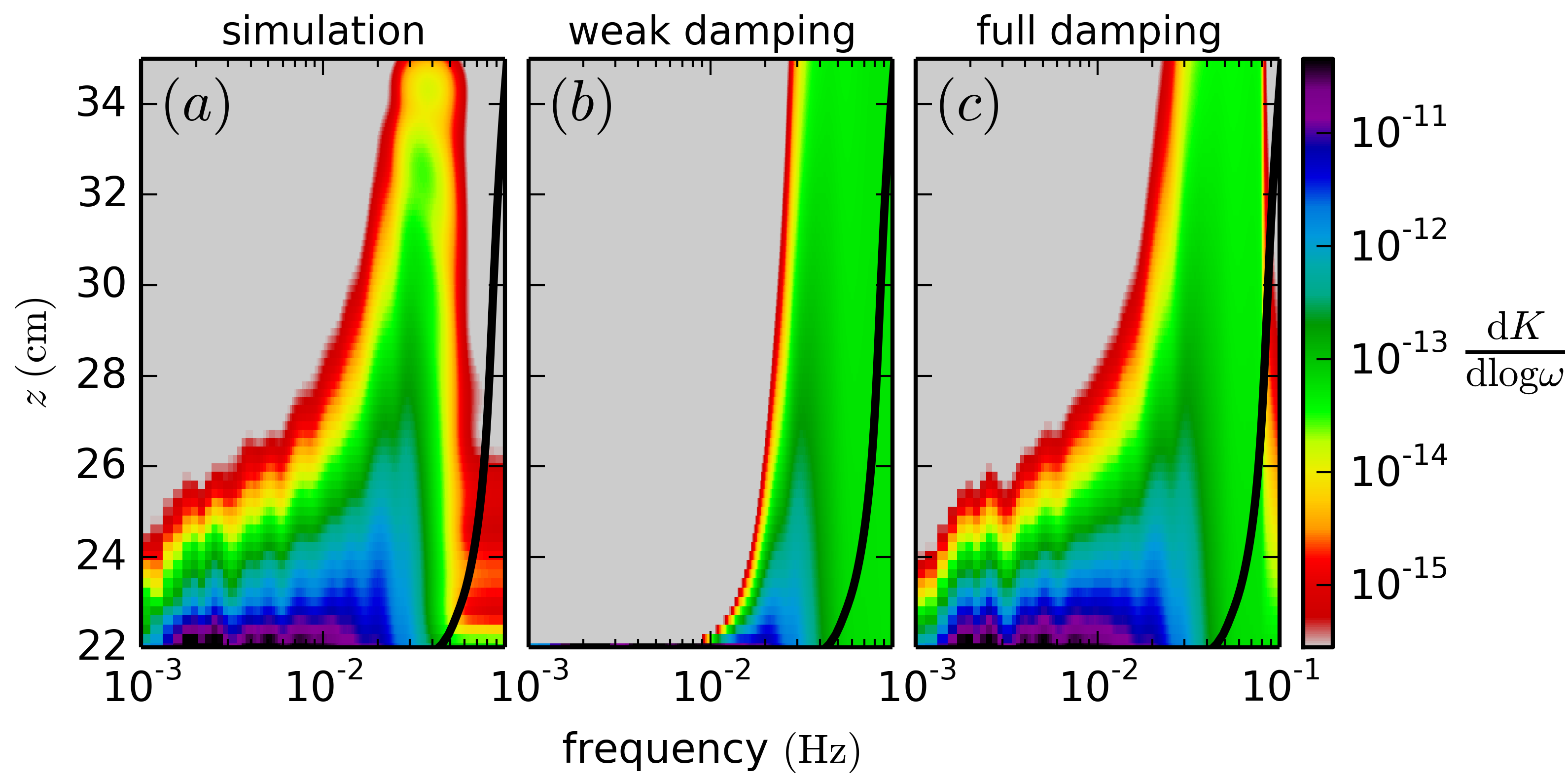}}
  \caption{(Color online) Spectrograms of kinetic energy in modes with $k_x=2\pi/0.1 \ {\rm m}^{-1}$ testing different linear damping formulas.  The black line shows the horizontally and temporally averaged buoyancy frequency profile.  Panel ($a$): the spectrogram for the simulation result. Panel ($b$): the spectrum from the simulation evaluated at $z=0.22 \ {\rm m}$ multiplied by the damping factor (equation~\ref{damping factor}), using the weak dissipation limit expression for $\ell_d^{-1}$ (equation~\ref{weak damping}). Panel ($c$): the spectrum from the simulation evaluated at $z=0.22 \ {\rm m}$ multiplied by the damping factor, using the full expression for $\ell_d^{-1}$ (equation~\ref{full damping}).  While the weak damping formula significantly overestimates the damping, the full damping formula accurately describes the damping in the simulation.}
\label{fig:damping}
\end{figure*}

In figure~\ref{fig:damping}, we compare these two damping rate calculations with the simulation.  Because the inverse damping length depends on the wavenumber, we first filter the simulation data to only include modes with $k_x=2\pi/0.1 \ {\rm m}^{-1}$.  Panel ($a$) shows the spectrogram of the kinetic energy in these modes in the simulation.  Note that there is a factor of $\sim 10$ less energy in this set of modes than in the full simulation (figure~\ref{fig:overview} ($a$)).  To simplify the calculation of the damping factor (equation~\ref{damping factor}), we approximate $N$ to be uniform and equal to $0.08 \ {\rm Hz}$.  In panel ($b$), we plot the energy as a function of height using equation~\ref{damping factor}, using the weak dissipation limit for $\ell_d^{-1}$ (equation~\ref{weak damping}), normalizing the wave energy at $z=0.22 \ {\rm m}$ using the simulation results \citep[as done in, e.g.,][]{ts07}.  This greatly overestimates the damping for short wavelength waves ($\omega<2\times 10^{-2} \ {\rm Hz}$) because these waves are not in the weak damping limit.  This is similar to the results of \citet{rm11,rogers13}, and \citet{alvan14}.  We repeat the calculation in panel ($c$), but now use the full expression for $\ell_d^{-1}$ (equation~\ref{full damping}).  When using the full expression without taking the weak dissipation limit, we find that linear theory correctly describes the damping in the stably stratified region.

\section{Bulk excitation: Reynolds stress forcing}\label{sec:source}

We now test two proposed models of internal wave excitation, first bulk excitation, then interface forcing (section~\ref{sec:interface}).  We use linear wave models and data from the full simulation to calculate the vertical velocity associated with the wave field.  We then compare the wave field in these simplified models with the wave field in the full simulation.

The bulk excitation mechanism is an application of the \citet{lighthill01} approach to calculating wave excitation.  We decompose the vertical displacement $\xi_z$ into a wave-like component and a convection-like component using the linear eigenfunctions.  The wave excitation is given by the projection of the Reynolds stress associated with convection onto the wave-like component by a source term $S$ \citep{lq13}.  This can be written as
\Beq\label{source equation}
\nabla^2&&\left(\partial_t - \nu\nabla^2\right) \partial_t\xi_z + N^2\partial_x^2 \xi_z = S \nonumber \\
&&\quad \quad = -\nabla^2 \left(\vec{u}\dot\grad u_z\right) + \partial_z\left[\left(\partial_{x_i} u_j\right)\left(\partial_{x_j} u_i\right)\right],
\Eeq
where repeated indices are summed over.  We have assumed here that the only important nonlinearity is the Reynolds stress and have neglected the $\vec{u}\dot\grad T$ nonlinearity.  Mixing length theory suggests the Reynolds stress term is the most important source term \citep[e.g.,][]{gk90}.  We also neglect thermal diffusivity because it has a smaller effect than viscosity.

To test the bulk excitation mechanism, we solve equation~\ref{source equation}, where the source term is calculated directly from the velocities in the full simulation.  This is possible by exploiting the flexibility of Dedalus.  We solve two problems simultaneously; the full problem described in section~\ref{sec:full}, as well as the linear forced problem described here.  At every time step, we first evolve the full problem forward. Then we calculate the source term using the velocities in the full problem.  With the new source term, we take time step in the linear forced problem, and repeat.

\citet{freund01} and \citet{boersma05} used a similar approach to study the generation of sound waves by a turbulent jet.  In \citet{freund01}, the linear model is used to study sound waves in the far field, which is too large to to be resolved by the jet simulation.  In contrast, \citet{boersma05} solves the incompressible equations, and uses a linear wave equation to estimate wave generation due to a source term.  Neither study is able to compare the predictions of the bulk excitation model with self-consistently generated waves within their simulations, and both limit their analysis to the far field.

We solve the linear problem on the same domain as the full problem, and use the same timestepping scheme.  We start the calculation at $t=34878 \ {\rm s}$.  The $N^2$ profile we use is a horizontally and time averaged profile, as in figure~\ref{fig:overview}($a$), where all negative values of $N^2$ are set to zero so that we can neglect exponentially growing convective modes.  The boundary conditions are $\xi_z=0$ on top and bottom, $\partial_t\partial_z\xi_z=\partial_x u=0$ on the top, and $\partial_t\partial_z^2\xi_z=\partial_x\partial_z u =0$ on the bottom, where $u$ is the horizontal velocity.  To make sure the source term arises only from the convection and not the existing waves in the simulation, we mask $S$ to include only the convection zone by multiplying it by $0.5 \left[1-\tanh((z-0.23)/0.01)\right]$.

\begin{figure*}
  \centerline{\includegraphics{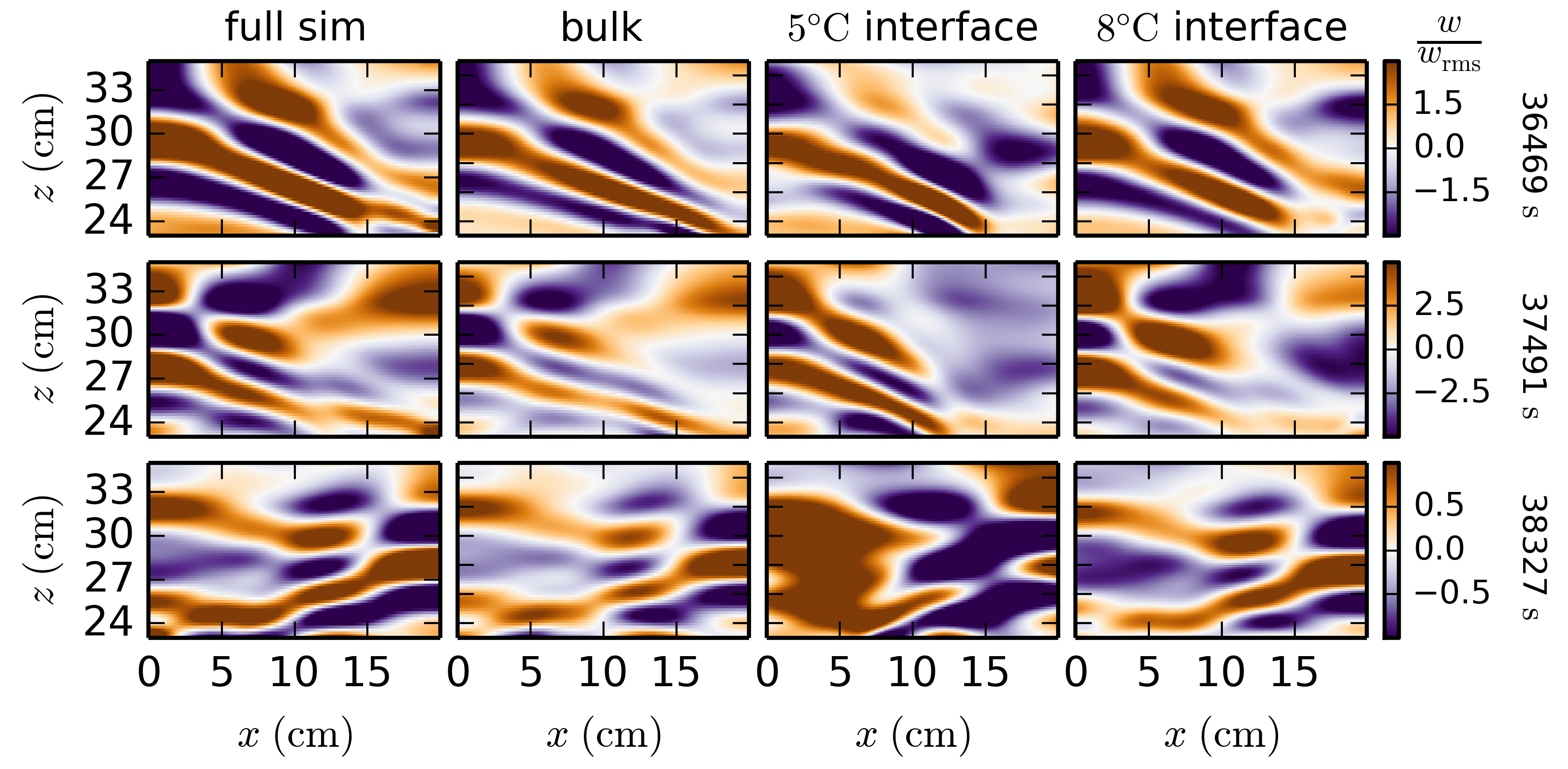}}
  \caption{(Color online) Snapshots of vertical velocity in the stably stratified region at three different times for the full simulation (first column), bulk excitation simulation (second column), and interface forcing simulations with a $5\degree{\rm C}$ interface (third column) and an $8\degree{\rm C}$ interface (fourth column).  The vertical velocities are normalized by the rms vertical velocity at each height.  Although there is quite good agreement between the full simulation and the bulk excitation simulation, the interface forcing simulations are not as accurate.}
\label{fig:flow}
\end{figure*}

In figure~\ref{fig:flow}, we compare the flow pattern between the full simulation and the bulk excitation simulation, as well as two interface forcing simulations described in section~\ref{sec:interface}.  Because the velocities decrease rapidly with height (see figure~\ref{fig:excitation}), we normalize the vertical velocity to its rms at each height.  At each of the three chosen times, there is very good agreement between the full simulation and the bulk excitation simulation.  Note that the second row corresponds to the same time as figure~\ref{fig:excitation}($d$).  The average amplitude for the bulk excitation simulation is smaller than the average amplitude of the full simulation by a factor of about $0.7$.  The agreement is quite impressive given the simplified physics in the bulk excitation simulation.  The remaining differences might be because we only included the Reynolds stress source term, and not a source term proportional to the $\vec{u}\dot\grad T$ nonlinearity.

\begin{figure*}
  \centerline{\includegraphics{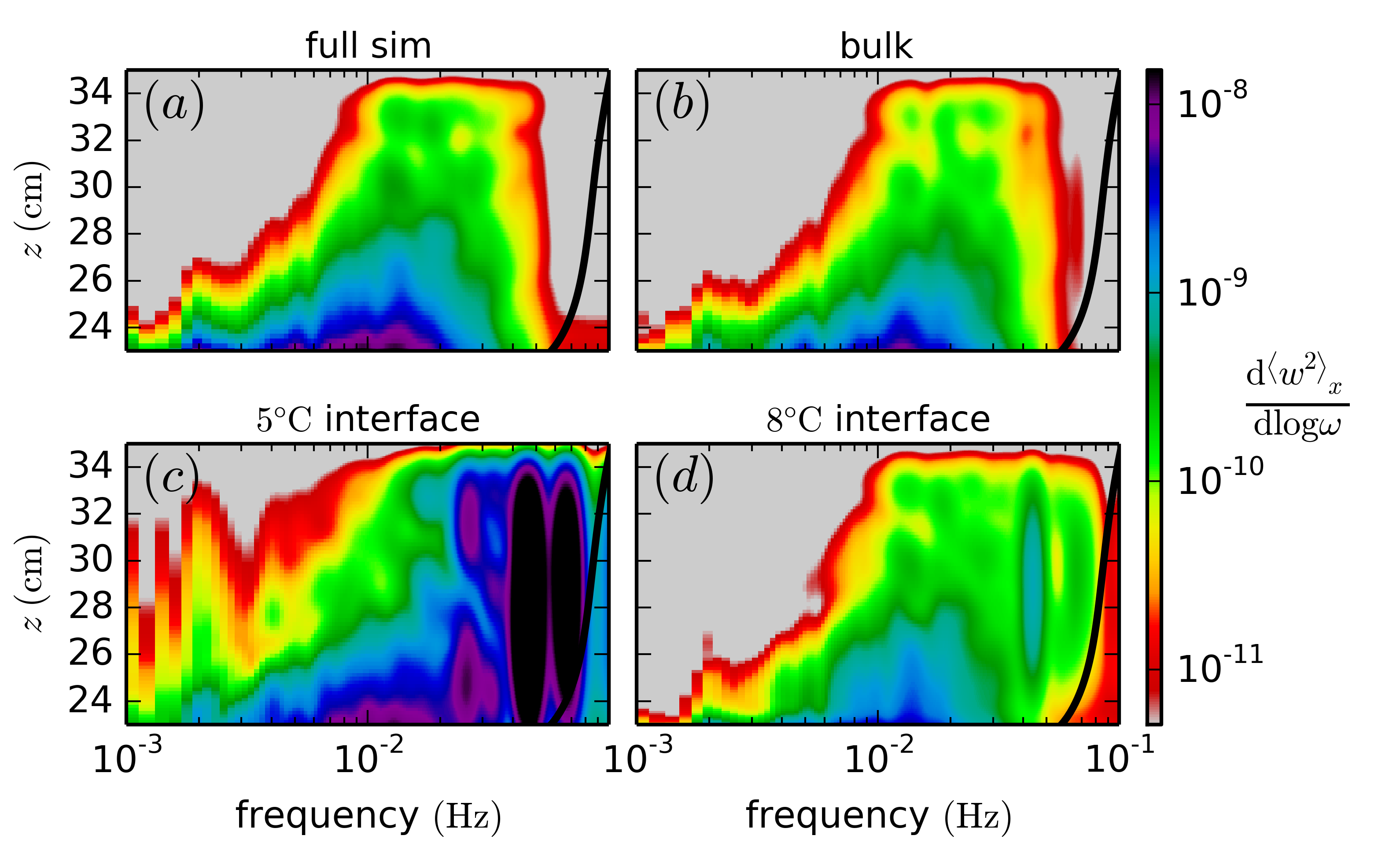}}
  \caption{(Color online) Spectrogram of vertical velocity squared in four different simulations.  The black line shows the horizontally and temporally averaged buoyancy frequency profile. Panel ($a$),~full simulation; panel ($b$),~bulk excitation simulation; interface forcing simulations with $5\degree{\rm C}$~interface~(panel $c$) and $8\degree{\rm C}$~interface~(panel $d$).  The bulk excitation simulation agrees very well with the full simulation.  The interface forcing simulations overestimate the excitation of high frequency waves.  This is because the interface forcing cannot detect sweeping motions along the interface, and thus treats the wave excitation as impulsive events which produce high frequency waves.}
\label{fig:comparison}
\end{figure*}

To compare quantitatively the two wave fields throughout the simulation, we plot the spectrogram for each simulation in figure~\ref{fig:comparison}.  Again, there is striking agreement between the full simulation and the bulk excitation simulation.  We also calculated the correlation of $w/w_{\rm rms}$ between the two simulations, from $35287 \ {\rm s}$ to $39121 \ {\rm s}$ and from $z=0.235 \ {\rm m}$ to $0.349 \ {\rm m}$.  The correlation between the full simulation and the bulk excitation simulation is 96\%.

\section{Interface forcing: Mechanical oscillator effect}\label{sec:interface}

In this section we present simulations which test interface forcing.  We again solve the linear IGW equation, but now force the system with a boundary condition instead of a source term.  The moving bottom boundary represents the movement of the interface between the convective and stably stratified regions in the full simulation.  This generates IGWs just as moving the surface of a drum generates sound waves.

First we calculate the interface position as a function of time using data from the full simulation.  We calculate two interfaces: a $5\degree {\rm C}$ interface and an $8\degree{\rm C}$ interface.  We use $5\degree{\rm C}$ to get very close to the density maximum ($4\degree{\rm C}$), without having to worry about large temperature fluctuations producing unphysical fluctuations in the interface position.  The $8\degree{\rm C}$ interface is used because a rising fluid element at $0\degree{\rm C}$ becomes neutrally buoyant when it reaches $8\degree{\rm C}$, assuming no conductive losses.  We calculate the interface position by, every two hundred time steps (i.e., one fifteenth of a buoyancy period)  in the full simulation, calculating the $z$ position at which $T=5\degree{\rm C}$ or $8\degree{\rm C}$ at each $x$ position.  The interface position at intermediate times is reconstructed via linear interpolation.  Calculating the interface position every one hundred time steps (and using linear interpolation for intermediate times) produced no visible changes in the simulation results.

The interface forcing simulations are run on a reduced domain, with $x$ varying from $0$ to $0.2 \ {\rm m}$, but $z$ varying from the mean interface position to $0.35 \ {\rm m}$.  The mean interface position is $0.2096 \ {\rm m}$ for the $5\degree{\rm C}$ interface and is $0.225 \ {\rm m}$ for the $8\degree{\rm C}$ interface.  The horizontal resolution in both cases is 512 Fourier modes and grid points (with no dealiasing), and the vertical resolution in both cases is 128 Chebyshev modes and grid points (with no dealiasing).  Doubling the vertical resolution of the simulation had no noticeable effect on the wave field.

More specifically, we solve the linear, homogeneous wave equation,
\Beq\label{interface equation}
\nabla^2 \left(\partial_t - \nu \nabla^2\right)\partial_t\xi_z + N^2\partial_x^2 \xi_z = 0,
\Eeq
where $N^2$ is the temporally and horizontally averaged buoyancy frequency, as in section~\ref{sec:source}.  Our boundary conditions are as follows.  On the bottom, we set $\xi_z=\delta z_{\rm int}(x,t)$, where $\delta z_{\rm int}(x,t)$ in the deviation of the interface position from its mean position at that time (i.e., $\left\langle \delta z_{\rm int}(x,t)\right\rangle_x = 0$).  This ensures there is zero mass flux into the domain.  The second bottom boundary condition is $\partial_t\partial_z^2\xi_z=\partial_x\partial_z u =0$.  We also tried imposing pressure continuity at the interface as a second boundary condition---this gives qualitatively similar results.  The top boundary conditions are $\xi_z=0$ and $\partial_t\partial_z\xi_z=\partial_x u=0$.

For timestepping, we use a first order forward Euler/backward Euler scheme.  Although the entire equation is linear, we treat the buoyancy term $N^2\partial_x^2\xi_z$ explicitly.  As mentioned above, the interface position is calculated every two hundred time steps in the full calculation---call this time difference $\Delta t_i$ (which varies, as the time step is set by the CFL condition).  The time step for these interface forcing calculations is $\Delta t_i/40$.  Thus, the average time step is five times larger than the average time step in the full simulation.  Using a smaller timestep of $\Delta t_i/80$ had no noticeable effect on the wave field.  The calculation starts at $t=34878 \ {\rm s}$.

Figure~\ref{fig:flow} shows three snapshots of the wave field for the interface forcing simulations using the $5\degree{\rm C}$ interface and the $8\degree{\rm C}$ interface.  Although the broad features of the wave field are present, the $5\degree{\rm C}$ interface simulation does not reproduce all the features of the full simulation (the difference is especially striking in the third snapshot at $t=38327 \ {\rm s}$).  However, the $8\degree{\rm C}$ interface simulation matches the full simulation fairly well.

Although it cannot be seen in the snapshots, a movie of the simulations (see supplementary materials) shows high frequency oscillations in both the $5\degree{\rm C}$ interface simulation and the $8\degree{\rm C}$ interface simulation.  This feature clearly is present in the spectrograms of the $5\degree{\rm C}$ interface simulation and the $8\degree{\rm C}$ interface simulation in figure~\ref{fig:comparison}.  Both show significant excess power at high frequency.

As for the bulk excitation simulation, we calculated the correlation of $w/w_{\rm rms}$ between the full simulation and each of the interface forcing simulations.  While the $8\degree{\rm C}$ interface simulation has a correlation of 81\%, the $5\degree{\rm C}$ interface simulation only has a correlation of 31\%.  Neither come close to matching the 96\% correlation of the bulk excitation simulation with the full simulation.

The interface forcing simulations over-excite high frequency waves because they treat the wave excitation as an impulsive process.  Large plumes can significantly deflect the interface on short timescales, which excites high frequency waves.  In the laboratory experiments of \citet{as10}, dense plumes produce these sorts of large deflections, generating high frequency waves.  \citet{michaelian02} and \citet{lebars15} also find that high frequency waves are generated when the changing thermal states early in their experiments produce vigorous plumes.  In the terminology of astrophysical penetrative convection, these would be termed ``penetrating'' plumes, rather than typical ``overshooting'' plumes \citep[e.g.,][]{bct02}.

However, most waves are not being generated by these particularly strong, ``penetrating'' plumes.  Instead, they are being generated by the sweeping motion of plumes below the interface, which preferentially generates waves at low frequencies and large horizontal wavelengths.  This is accurately captured by the bulk excitation calculation, but not in the interface forcing calculations.  If something prevented motions in the convection zone near the interface, there could be waves generated by stresses within the convection zone, even if there are no accompanying interface motions.  Because the interface forcing calculations do not know about these sweeping motions, they treat the plumes as an impulsive forcing, generating too many high frequency waves.

Using a higher interface (the $8\degree{\rm C}$ interface) improves the results as the dynamics are better modeled as linear.  This is an important check on our numerical implementation of the interface forcing simulations.  An interface which is far away from the convection should faithfully reproduce the wave field, but this does not give insight into the wave excitation process since the ``interface'' motions are dominated by the waves themselves rather than the convection.  Instead, this exposes a conceptual problem with the interface forcing picture of wave excitation.

\section{Conclusions}\label{sec:conclusion}

In this paper we present simulations of convective excitation of internal gravity waves in a fluid with a water-like equation of state.  Cold, buoyant plumes near $0\degree{\rm C}$ detach from an unstable boundary layer at the bottom of the domain and are advected by two convective cells to the top of the convective region.  As the plumes approach the interface between the convective and stably stratified regions, they generate internal gravity waves (figure~\ref{fig:excitation}).  Although the wave excitation is dominated by low frequency waves matching the convective turnover frequency, these waves are quickly damped as they propagate upwards.  Only high frequency waves reach the top of the domain (figure~\ref{fig:overview}).  Estimates of the wave flux are reasonably consistent with the theoretical estimate of $\mathcal{M} u_c^3$, where $\mathcal{M}$ is the ratio of convection frequency to buoyancy frequency, and $u_c$ is a characteristic convection velocity.

We find that the wave damping with height is in agreement with analytic theory, provided one uses the full expression for the damping length (equation~\ref{full damping}).  Because geophysical and astrophysical waves excited by convection are typically weakly damped by diffusive processes, previous analyses of simulations \citep[e.g.,][]{rm11,rogers13,alvan14} have used the quasi-adiabatic, or weak dissipation limit (equation~\ref{weak damping}).  However, simulations and laboratory experiments have much larger diffusivities than natural systems, and we find that the waves generated in our simulation are mostly not in the weak dissipation limit.  Using the full damping expression, we find that low frequency waves have inverse damping lengths proportional to $\omega^{-1/4}$, rather than $\omega^{-4}$ in the weak dissipation limit.  This implies that the weak dissipation expression overestimates the wave damping, consistent with previous claims about the failure of linear damping to explain the damping in simulations of convective excited waves.  However, we find that the full linear damping rate is able to correctly reproduce the damping in the simulation (figure~\ref{fig:damping}).

In addition to this full simulation, we also present three simulations of the simulation.  The first tests bulk excitation, where waves are excited by Reynolds stresses within the convective region (section~\ref{sec:source}).  We solve the linear wave equation forced by a Reynolds stress source term calculated at each position and time from the convection zone of the full simulation.  This accurately reproduces the wave field and spectrum (figures~\ref{fig:flow} \& \ref{fig:comparison}); the wave field correlation between the bulk excitation and full simulations is 96\%.  The wave amplitude is underestimated by a factor of $0.7$---this may be because we neglect other nonlinear source terms like $\vec{u}\dot\grad T$ in our calculation.

We also run two simulations testing interface forcing (section~\ref{sec:interface}).  In these simulations, we solve the homogeneous linear wave equation, but force the system with the motions of either the $5\degree{\rm C}$ isotherm or the $8\degree{\rm C}$ isotherm at the bottom of the stably stratified region.  The movement of these isotherms is analogous to the oscillations of the surface of a drum pummeled by the mallets of convection.  However, this neglects the sweeping motion of plumes below the interface.  These simulations thus misinterpret the convective excitation as being due to impulsive, penetrating plumes, which over-excite high frequency waves (figure~\ref{fig:comparison}).  The inability to detect the horizontal motion of sweeping plumes is a fundamental problem of the interface forcing heuristic.  In contrast to the 96\% correlation between the wave fields in the bulk excitation and full simulations, the $5\degree{\rm C}$ interface forcing and the $8\degree{\rm C}$ interface forcing simulations have a correlation with the full simulation of only 31\% and 81\%, respectively.

All four simulations presented in this paper exploit the flexibility of Dedalus.  Modifying the classic Boussinesq system to accommodate the nonlinear equation of state of water amounts to changing one line of the simulation script.  Furthermore, being able to solve different equations easily makes it practical to test different heuristics by solving model equations using simulation data as input.  This represents a novel and fruitful analysis technique.

Although the results presented here are for two dimensional simulations using the unique equation of state of water near $4\degree{\rm C}$, we expect similar conclusions to hold in three dimensional, and for fluids with less exotic thermodynamic properties.  In future work, we will run similar calculations in three dimensions at high Rayleigh number, including more physics, such as density stratification and rotation.

\section*{Acknowledgments}
We thank Tami Rogers, Sacha Brun, Stephane Mathis, Jim Fuller and Bruce Sutherland for helpful discussions.  DL is supported by a Hertz Foundation Fellowship, the National Science Foundation Graduate Research Fellowship under Grant No. DGE 1106400, a Kavli Institute for Theoretical Physics Graduate Student Fellowship, and partially by the Schneider Chair in Physics to Eliot Quataert.
MLB acknowledges support from the Marie Curie Actions of the European Commission (FP7-PEOPLE-2011-IOF). MLB and DL thank the Labex program MEC (ANR-11-LABX-0092) for  supporting DL visit in Marseilles. KJB is supported by a MIT Kavli Institute Graduate Fellowship and a National Science Foundation Graduate Research Fellowship under Grant No. 1122374. GMV is generously supported by the Australia Research Council, project number DE140101960. BPB acknowledges support from NSF Astronomy and Astrophysics postdoctoral fellowship AST 09-02004 and a KITP postdoctoral fellowship (Grant No. NSF PHY11-25915).  This work was supported in part by NASA ATP Grant 12- ATP12-0183, by NASA grant NNX10AJ96G, by the David and Lucile Packard Foundation, and by a Simons Investigator Award to EQ from the Simons Foundation. JSO acknowledges support from NSF Grant AST10-09802. Part of this work was completed at the Kavli Institute of Theoretical Physics programs on Wave-Mean Flow Interaction, and Star Formation (Grant No. NSF PHY11-25915).

\bibliography{water}

\begin{thebibliography}{35}%
\makeatletter
\providecommand \@ifxundefined [1]{%
 \@ifx{#1\undefined}
}%
\providecommand \@ifnum [1]{%
 \ifnum #1\expandafter \@firstoftwo
 \else \expandafter \@secondoftwo
 \fi
}%
\providecommand \@ifx [1]{%
 \ifx #1\expandafter \@firstoftwo
 \else \expandafter \@secondoftwo
 \fi
}%
\providecommand \natexlab [1]{#1}%
\providecommand \enquote  [1]{``#1''}%
\providecommand \bibnamefont  [1]{#1}%
\providecommand \bibfnamefont [1]{#1}%
\providecommand \citenamefont [1]{#1}%
\providecommand \href@noop [0]{\@secondoftwo}%
\providecommand \href [0]{\begingroup \@sanitize@url \@href}%
\providecommand \@href[1]{\@@startlink{#1}\@@href}%
\providecommand \@@href[1]{\endgroup#1\@@endlink}%
\providecommand \@sanitize@url [0]{\catcode `\\12\catcode `\$12\catcode
  `\&12\catcode `\#12\catcode `\^12\catcode `\_12\catcode `\%12\relax}%
\providecommand \@@startlink[1]{}%
\providecommand \@@endlink[0]{}%
\providecommand \url  [0]{\begingroup\@sanitize@url \@url }%
\providecommand \@url [1]{\endgroup\@href {#1}{\urlprefix }}%
\providecommand \urlprefix  [0]{URL }%
\providecommand \Eprint [0]{\href }%
\providecommand \doibase [0]{http://dx.doi.org/}%
\providecommand \selectlanguage [0]{\@gobble}%
\providecommand \bibinfo  [0]{\@secondoftwo}%
\providecommand \bibfield  [0]{\@secondoftwo}%
\providecommand \translation [1]{[#1]}%
\providecommand \BibitemOpen [0]{}%
\providecommand \bibitemStop [0]{}%
\providecommand \bibitemNoStop [0]{.\EOS\space}%
\providecommand \EOS [0]{\spacefactor3000\relax}%
\providecommand \BibitemShut  [1]{\csname bibitem#1\endcsname}%
\let\auto@bib@innerbib\@empty
\bibitem [{\citenamefont {{Baldwin}}\ \emph {et~al.}(2001)\citenamefont
  {{Baldwin}}, \citenamefont {{Gray}}, \citenamefont {{Dunkerton}},
  \citenamefont {{Hamilton}}, \citenamefont {{Haynes}}, \citenamefont
  {{Randel}}, \citenamefont {{Holton}}, \citenamefont {{Alexander}},
  \citenamefont {{Hirota}}, \citenamefont {{Horinouchi}}, \citenamefont
  {{Jones}}, \citenamefont {{Kinnersley}}, \citenamefont {{Marquardt}},
  \citenamefont {{Sato}},\ and\ \citenamefont {{Takahashi}}}]{baldwin01}%
  \BibitemOpen
  \bibfield  {author} {\bibinfo {author} {\bibfnamefont {M.~P.}\ \bibnamefont
  {{Baldwin}}}, \bibinfo {author} {\bibfnamefont {L.~J.}\ \bibnamefont
  {{Gray}}}, \bibinfo {author} {\bibfnamefont {T.~J.}\ \bibnamefont
  {{Dunkerton}}}, \bibinfo {author} {\bibfnamefont {K.}~\bibnamefont
  {{Hamilton}}}, \bibinfo {author} {\bibfnamefont {P.~H.}\ \bibnamefont
  {{Haynes}}}, \bibinfo {author} {\bibfnamefont {W.~J.}\ \bibnamefont
  {{Randel}}}, \bibinfo {author} {\bibfnamefont {J.~R.}\ \bibnamefont
  {{Holton}}}, \bibinfo {author} {\bibfnamefont {M.~J.}\ \bibnamefont
  {{Alexander}}}, \bibinfo {author} {\bibfnamefont {I.}~\bibnamefont
  {{Hirota}}}, \bibinfo {author} {\bibfnamefont {T.}~\bibnamefont
  {{Horinouchi}}}, \bibinfo {author} {\bibfnamefont {D.~B.~A.}\ \bibnamefont
  {{Jones}}}, \bibinfo {author} {\bibfnamefont {J.~S.}\ \bibnamefont
  {{Kinnersley}}}, \bibinfo {author} {\bibfnamefont {C.}~\bibnamefont
  {{Marquardt}}}, \bibinfo {author} {\bibfnamefont {K.}~\bibnamefont {{Sato}}},
  \ and\ \bibinfo {author} {\bibfnamefont {M.}~\bibnamefont {{Takahashi}}},\
  }\href {\doibase 10.1029/1999RG000073} {\bibfield  {journal} {\bibinfo
  {journal} {Reviews of Geophysics}\ }\textbf {\bibinfo {volume} {39}},\
  \bibinfo {pages} {179} (\bibinfo {year} {2001})}\BibitemShut {NoStop}%
\bibitem [{\citenamefont {{Talon}}\ \emph {et~al.}(2002)\citenamefont
  {{Talon}}, \citenamefont {{Kumar}},\ and\ \citenamefont {{Zahn}}}]{talon02}%
  \BibitemOpen
  \bibfield  {author} {\bibinfo {author} {\bibfnamefont {S.}~\bibnamefont
  {{Talon}}}, \bibinfo {author} {\bibfnamefont {P.}~\bibnamefont {{Kumar}}}, \
  and\ \bibinfo {author} {\bibfnamefont {J.-P.}\ \bibnamefont {{Zahn}}},\
  }\href {\doibase 10.1086/342526} {\bibfield  {journal} {\bibinfo  {journal}
  {\apjl}\ }\textbf {\bibinfo {volume} {574}},\ \bibinfo {pages} {L175}
  (\bibinfo {year} {2002})},\ \Eprint {http://arxiv.org/abs/astro-ph/0206479}
  {astro-ph/0206479} \BibitemShut {NoStop}%
\bibitem [{\citenamefont {{Talon}}\ and\ \citenamefont
  {{Charbonnel}}(2005)}]{tc05}%
  \BibitemOpen
  \bibfield  {author} {\bibinfo {author} {\bibfnamefont {S.}~\bibnamefont
  {{Talon}}}\ and\ \bibinfo {author} {\bibfnamefont {C.}~\bibnamefont
  {{Charbonnel}}},\ }\href {\doibase 10.1051/0004-6361:20053020} {\bibfield
  {journal} {\bibinfo  {journal} {\aap}\ }\textbf {\bibinfo {volume} {440}},\
  \bibinfo {pages} {981} (\bibinfo {year} {2005})},\ \Eprint
  {http://arxiv.org/abs/astro-ph/0505229} {astro-ph/0505229} \BibitemShut
  {NoStop}%
\bibitem [{\citenamefont {{Quataert}}\ and\ \citenamefont
  {{Shiode}}(2012)}]{qs12}%
  \BibitemOpen
  \bibfield  {author} {\bibinfo {author} {\bibfnamefont {E.}~\bibnamefont
  {{Quataert}}}\ and\ \bibinfo {author} {\bibfnamefont {J.}~\bibnamefont
  {{Shiode}}},\ }\href {\doibase 10.1111/j.1745-3933.2012.01264.x} {\bibfield
  {journal} {\bibinfo  {journal} {\mnras}\ }\textbf {\bibinfo {volume} {423}},\
  \bibinfo {pages} {L92} (\bibinfo {year} {2012})},\ \Eprint
  {http://arxiv.org/abs/1202.5036} {arXiv:1202.5036 [astro-ph.SR]} \BibitemShut
  {NoStop}%
\bibitem [{\citenamefont {{Shiode}}\ and\ \citenamefont
  {{Quataert}}(2014)}]{sq14}%
  \BibitemOpen
  \bibfield  {author} {\bibinfo {author} {\bibfnamefont {J.~H.}\ \bibnamefont
  {{Shiode}}}\ and\ \bibinfo {author} {\bibfnamefont {E.}~\bibnamefont
  {{Quataert}}},\ }\href {\doibase 10.1088/0004-637X/780/1/96} {\bibfield
  {journal} {\bibinfo  {journal} {\apj}\ }\textbf {\bibinfo {volume} {780}},\
  \bibinfo {eid} {96} (\bibinfo {year} {2014})},\ \Eprint
  {http://arxiv.org/abs/1308.5978} {arXiv:1308.5978 [astro-ph.SR]} \BibitemShut
  {NoStop}%
\bibitem [{\citenamefont {{Fuller}}\ \emph {et~al.}(2014)\citenamefont
  {{Fuller}}, \citenamefont {{Lecoanet}}, \citenamefont {{Cantiello}},\ and\
  \citenamefont {{Brown}}}]{fuller14}%
  \BibitemOpen
  \bibfield  {author} {\bibinfo {author} {\bibfnamefont {J.}~\bibnamefont
  {{Fuller}}}, \bibinfo {author} {\bibfnamefont {D.}~\bibnamefont
  {{Lecoanet}}}, \bibinfo {author} {\bibfnamefont {M.}~\bibnamefont
  {{Cantiello}}}, \ and\ \bibinfo {author} {\bibfnamefont {B.~P.}\ \bibnamefont
  {{Brown}}},\ }\href@noop {} {\enquote {\bibinfo {title} {{Angular Momentum
  Transport via Internal Gravity Waves in Evolving Stars}},}\ } (\bibinfo
  {year} {2014}),\ \bibinfo {note} {{Submitted to ApJ}}\BibitemShut {NoStop}%
\bibitem [{\citenamefont {{Goodman}}\ and\ \citenamefont
  {{Lackner}}(2009)}]{gl09}%
  \BibitemOpen
  \bibfield  {author} {\bibinfo {author} {\bibfnamefont {J.}~\bibnamefont
  {{Goodman}}}\ and\ \bibinfo {author} {\bibfnamefont {C.}~\bibnamefont
  {{Lackner}}},\ }\href {\doibase 10.1088/0004-637X/696/2/2054} {\bibfield
  {journal} {\bibinfo  {journal} {\apj}\ }\textbf {\bibinfo {volume} {696}},\
  \bibinfo {pages} {2054} (\bibinfo {year} {2009})},\ \Eprint
  {http://arxiv.org/abs/0812.1028} {arXiv:0812.1028} \BibitemShut {NoStop}%
\bibitem [{\citenamefont {{Fritts}}\ and\ \citenamefont
  {{Alexander}}(2003)}]{fa03}%
  \BibitemOpen
  \bibfield  {author} {\bibinfo {author} {\bibfnamefont {D.~C.}\ \bibnamefont
  {{Fritts}}}\ and\ \bibinfo {author} {\bibfnamefont {M.~J.}\ \bibnamefont
  {{Alexander}}},\ }\href {\doibase 10.1029/2001RG000106} {\bibfield  {journal}
  {\bibinfo  {journal} {Reviews of Geophysics}\ }\textbf {\bibinfo {volume}
  {41}},\ \bibinfo {eid} {1003} (\bibinfo {year} {2003})}\BibitemShut {NoStop}%
\bibitem [{\citenamefont {{Ansong}}\ and\ \citenamefont
  {{Sutherland}}(2010)}]{as10}%
  \BibitemOpen
  \bibfield  {author} {\bibinfo {author} {\bibfnamefont {J.~K.}\ \bibnamefont
  {{Ansong}}}\ and\ \bibinfo {author} {\bibfnamefont {B.~R.}\ \bibnamefont
  {{Sutherland}}},\ }\href {\doibase 10.1017/S0022112009993193} {\bibfield
  {journal} {\bibinfo  {journal} {Journal of Fluid Mechanics}\ }\textbf
  {\bibinfo {volume} {648}},\ \bibinfo {pages} {405} (\bibinfo {year}
  {2010})}\BibitemShut {NoStop}%
\bibitem [{\citenamefont {{Clark}}\ \emph {et~al.}(1986)\citenamefont
  {{Clark}}, \citenamefont {{Hauf}},\ and\ \citenamefont
  {{Kuettner}}}]{clark86}%
  \BibitemOpen
  \bibfield  {author} {\bibinfo {author} {\bibfnamefont {T.~L.}\ \bibnamefont
  {{Clark}}}, \bibinfo {author} {\bibfnamefont {T.}~\bibnamefont {{Hauf}}}, \
  and\ \bibinfo {author} {\bibfnamefont {J.~P.}\ \bibnamefont {{Kuettner}}},\
  }\href {\doibase 10.1002/qj.49711247402} {\bibfield  {journal} {\bibinfo
  {journal} {Quarterly Journal of the Royal Meteorological Society}\ }\textbf
  {\bibinfo {volume} {112}},\ \bibinfo {pages} {899} (\bibinfo {year}
  {1986})}\BibitemShut {NoStop}%
\bibitem [{\citenamefont {{Fovell}}\ \emph {et~al.}(1992)\citenamefont
  {{Fovell}}, \citenamefont {{Durran}},\ and\ \citenamefont
  {{Holton}}}]{fovell92}%
  \BibitemOpen
  \bibfield  {author} {\bibinfo {author} {\bibfnamefont {R.}~\bibnamefont
  {{Fovell}}}, \bibinfo {author} {\bibfnamefont {D.}~\bibnamefont {{Durran}}},
  \ and\ \bibinfo {author} {\bibfnamefont {J.~R.}\ \bibnamefont {{Holton}}},\
  }\href {\doibase 10.1175/1520-0469(1992)049<1427:NSOCGS>2.0.CO;2} {\bibfield
  {journal} {\bibinfo  {journal} {Journal of Atmospheric Sciences}\ }\textbf
  {\bibinfo {volume} {49}},\ \bibinfo {pages} {1427} (\bibinfo {year}
  {1992})}\BibitemShut {NoStop}%
\bibitem [{\citenamefont {{Goldreich}}\ and\ \citenamefont
  {{Kumar}}(1990)}]{gk90}%
  \BibitemOpen
  \bibfield  {author} {\bibinfo {author} {\bibfnamefont {P.}~\bibnamefont
  {{Goldreich}}}\ and\ \bibinfo {author} {\bibfnamefont {P.}~\bibnamefont
  {{Kumar}}},\ }\href {\doibase 10.1086/169376} {\bibfield  {journal} {\bibinfo
   {journal} {\apj}\ }\textbf {\bibinfo {volume} {363}},\ \bibinfo {pages}
  {694} (\bibinfo {year} {1990})}\BibitemShut {NoStop}%
\bibitem [{\citenamefont {{Lighthill}}(2001)}]{lighthill01}%
  \BibitemOpen
  \bibfield  {author} {\bibinfo {author} {\bibfnamefont {J.}~\bibnamefont
  {{Lighthill}}},\ }\href@noop {} {\emph {\bibinfo {title} {Waves in Fluids, by
  James Lighthill, pp.~520.~ISBN 0521010454.~Cambridge, UK: Cambridge
  University Press, December 2001.}}}\ (\bibinfo {year} {2001})\BibitemShut
  {NoStop}%
\bibitem [{\citenamefont {{Lesshafft}}\ \emph {et~al.}(2010)\citenamefont
  {{Lesshafft}}, \citenamefont {{Huerre}},\ and\ \citenamefont
  {{Sagaut}}}]{lesshafft10}%
  \BibitemOpen
  \bibfield  {author} {\bibinfo {author} {\bibfnamefont {L.}~\bibnamefont
  {{Lesshafft}}}, \bibinfo {author} {\bibfnamefont {P.}~\bibnamefont
  {{Huerre}}}, \ and\ \bibinfo {author} {\bibfnamefont {P.}~\bibnamefont
  {{Sagaut}}},\ }\href {\doibase 10.1017/S0022112009993612} {\bibfield
  {journal} {\bibinfo  {journal} {Journal of Fluid Mechanics}\ }\textbf
  {\bibinfo {volume} {647}},\ \bibinfo {pages} {473} (\bibinfo {year}
  {2010})}\BibitemShut {NoStop}%
\bibitem [{\citenamefont {{Townsend}}(1964)}]{townsend64}%
  \BibitemOpen
  \bibfield  {author} {\bibinfo {author} {\bibfnamefont {A.~A.}\ \bibnamefont
  {{Townsend}}},\ }\href {\doibase 10.1002/qj.49709038503} {\bibfield
  {journal} {\bibinfo  {journal} {Quarterly Journal of the Royal Meteorological
  Society}\ }\textbf {\bibinfo {volume} {90}},\ \bibinfo {pages} {248}
  (\bibinfo {year} {1964})}\BibitemShut {NoStop}%
\bibitem [{\citenamefont {{Townsend}}(1966)}]{townsend66}%
  \BibitemOpen
  \bibfield  {author} {\bibinfo {author} {\bibfnamefont {A.~A.}\ \bibnamefont
  {{Townsend}}},\ }\href {\doibase 10.1017/S0022112066000661} {\bibfield
  {journal} {\bibinfo  {journal} {Journal of Fluid Mechanics}\ }\textbf
  {\bibinfo {volume} {24}},\ \bibinfo {pages} {307} (\bibinfo {year}
  {1966})}\BibitemShut {NoStop}%
\bibitem [{\citenamefont {{Veronis}}(1963)}]{veronis63}%
  \BibitemOpen
  \bibfield  {author} {\bibinfo {author} {\bibfnamefont {G.}~\bibnamefont
  {{Veronis}}},\ }\href {\doibase 10.1086/147538} {\bibfield  {journal}
  {\bibinfo  {journal} {\apj}\ }\textbf {\bibinfo {volume} {137}},\ \bibinfo
  {pages} {641} (\bibinfo {year} {1963})}\BibitemShut {NoStop}%
\bibitem [{\citenamefont {{Brummell}}(1993)}]{brummell93}%
  \BibitemOpen
  \bibfield  {author} {\bibinfo {author} {\bibfnamefont {N.~H.}\ \bibnamefont
  {{Brummell}}},\ }\href {\doibase 10.1080/03091929308203564} {\bibfield
  {journal} {\bibinfo  {journal} {Geophysical and Astrophysical Fluid
  Dynamics}\ }\textbf {\bibinfo {volume} {68}},\ \bibinfo {pages} {115}
  (\bibinfo {year} {1993})}\BibitemShut {NoStop}%
\bibitem [{\citenamefont {{Moore}}\ and\ \citenamefont {{Weiss}}(1973)}]{mw73}%
  \BibitemOpen
  \bibfield  {author} {\bibinfo {author} {\bibfnamefont {D.~R.}\ \bibnamefont
  {{Moore}}}\ and\ \bibinfo {author} {\bibfnamefont {N.~O.}\ \bibnamefont
  {{Weiss}}},\ }\href {\doibase 10.1017/S0022112073000868} {\bibfield
  {journal} {\bibinfo  {journal} {Journal of Fluid Mechanics}\ }\textbf
  {\bibinfo {volume} {61}},\ \bibinfo {pages} {553} (\bibinfo {year}
  {1973})}\BibitemShut {NoStop}%
\bibitem [{\citenamefont {{Rogers}}\ \emph {et~al.}(2013)\citenamefont
  {{Rogers}}, \citenamefont {{Lin}}, \citenamefont {{McElwaine}},\ and\
  \citenamefont {{Lau}}}]{rogers13}%
  \BibitemOpen
  \bibfield  {author} {\bibinfo {author} {\bibfnamefont {T.~M.}\ \bibnamefont
  {{Rogers}}}, \bibinfo {author} {\bibfnamefont {D.~N.~C.}\ \bibnamefont
  {{Lin}}}, \bibinfo {author} {\bibfnamefont {J.~N.}\ \bibnamefont
  {{McElwaine}}}, \ and\ \bibinfo {author} {\bibfnamefont {H.~H.~B.}\
  \bibnamefont {{Lau}}},\ }\href {\doibase 10.1088/0004-637X/772/1/21}
  {\bibfield  {journal} {\bibinfo  {journal} {\apj}\ }\textbf {\bibinfo
  {volume} {772}},\ \bibinfo {eid} {21} (\bibinfo {year} {2013})},\ \Eprint
  {http://arxiv.org/abs/1306.3262} {arXiv:1306.3262 [astro-ph.SR]} \BibitemShut
  {NoStop}%
\bibitem [{\citenamefont {{Alvan}}\ \emph {et~al.}(2014)\citenamefont
  {{Alvan}}, \citenamefont {{Brun}},\ and\ \citenamefont {{Mathis}}}]{alvan14}%
  \BibitemOpen
  \bibfield  {author} {\bibinfo {author} {\bibfnamefont {L.}~\bibnamefont
  {{Alvan}}}, \bibinfo {author} {\bibfnamefont {A.~S.}\ \bibnamefont {{Brun}}},
  \ and\ \bibinfo {author} {\bibfnamefont {S.}~\bibnamefont {{Mathis}}},\
  }\href {\doibase 10.1051/0004-6361/201323253} {\bibfield  {journal} {\bibinfo
   {journal} {\aap}\ }\textbf {\bibinfo {volume} {565}},\ \bibinfo {eid} {A42}
  (\bibinfo {year} {2014})},\ \Eprint {http://arxiv.org/abs/1403.4052}
  {arXiv:1403.4052 [astro-ph.SR]} \BibitemShut {NoStop}%
\bibitem [{\citenamefont {{Belkacem}}\ \emph {et~al.}(2009)\citenamefont
  {{Belkacem}}, \citenamefont {{Samadi}}, \citenamefont {{Goupil}},
  \citenamefont {{Dupret}}, \citenamefont {{Brun}},\ and\ \citenamefont
  {{Baudin}}}]{belkacem09}%
  \BibitemOpen
  \bibfield  {author} {\bibinfo {author} {\bibfnamefont {K.}~\bibnamefont
  {{Belkacem}}}, \bibinfo {author} {\bibfnamefont {R.}~\bibnamefont
  {{Samadi}}}, \bibinfo {author} {\bibfnamefont {M.~J.}\ \bibnamefont
  {{Goupil}}}, \bibinfo {author} {\bibfnamefont {M.~A.}\ \bibnamefont
  {{Dupret}}}, \bibinfo {author} {\bibfnamefont {A.~S.}\ \bibnamefont
  {{Brun}}}, \ and\ \bibinfo {author} {\bibfnamefont {F.}~\bibnamefont
  {{Baudin}}},\ }\href {\doibase 10.1051/0004-6361:200810827} {\bibfield
  {journal} {\bibinfo  {journal} {\aap}\ }\textbf {\bibinfo {volume} {494}},\
  \bibinfo {pages} {191} (\bibinfo {year} {2009})},\ \Eprint
  {http://arxiv.org/abs/0810.0602} {arXiv:0810.0602} \BibitemShut {NoStop}%
\bibitem [{\citenamefont {{Showman}}\ and\ \citenamefont
  {{Kaspi}}(2013)}]{sk13}%
  \BibitemOpen
  \bibfield  {author} {\bibinfo {author} {\bibfnamefont {A.~P.}\ \bibnamefont
  {{Showman}}}\ and\ \bibinfo {author} {\bibfnamefont {Y.}~\bibnamefont
  {{Kaspi}}},\ }\href {\doibase 10.1088/0004-637X/776/2/85} {\bibfield
  {journal} {\bibinfo  {journal} {\apj}\ }\textbf {\bibinfo {volume} {776}},\
  \bibinfo {eid} {85} (\bibinfo {year} {2013})},\ \Eprint
  {http://arxiv.org/abs/1210.7573} {arXiv:1210.7573 [astro-ph.EP]} \BibitemShut
  {NoStop}%
\bibitem [{\citenamefont {{Burns}}\ \emph {et~al.}(2015)\citenamefont
  {{Burns}}, \citenamefont {{Vasil}}, \citenamefont {{Oishi}}, \citenamefont
  {{Lecoanet}}, \citenamefont {{Brown}},\ and\ \citenamefont
  {{Quataert}}}]{burns15}%
  \BibitemOpen
  \bibfield  {author} {\bibinfo {author} {\bibfnamefont {K.~J.}\ \bibnamefont
  {{Burns}}}, \bibinfo {author} {\bibfnamefont {G.~M.}\ \bibnamefont
  {{Vasil}}}, \bibinfo {author} {\bibfnamefont {J.~S.}\ \bibnamefont
  {{Oishi}}}, \bibinfo {author} {\bibfnamefont {D.}~\bibnamefont {{Lecoanet}}},
  \bibinfo {author} {\bibfnamefont {B.~P.}\ \bibnamefont {{Brown}}}, \ and\
  \bibinfo {author} {\bibfnamefont {E.}~\bibnamefont {{Quataert}}},\
  }\href@noop {} {\enquote {\bibinfo {title} {{Dedalus: A Flexible
  Pseudo-Spectral Framework for Solving Partial Differential Equations}},}\ }
  (\bibinfo {year} {2015}),\ \bibinfo {note} {{In preparation}}\BibitemShut
  {NoStop}%
\bibitem [{\citenamefont {{Le Bars}}\ \emph {et~al.}(2015)\citenamefont {{Le
  Bars}}, \citenamefont {{Lecoanet}}, \citenamefont {{Aurnou}}, \citenamefont
  {{Perrard}}, \citenamefont {{Ribeiro}}, \citenamefont {{Rodet}},\ and\
  \citenamefont {{Le Gal}}}]{lebars15}%
  \BibitemOpen
  \bibfield  {author} {\bibinfo {author} {\bibfnamefont {M.}~\bibnamefont {{Le
  Bars}}}, \bibinfo {author} {\bibfnamefont {D.}~\bibnamefont {{Lecoanet}}},
  \bibinfo {author} {\bibfnamefont {J.~M.}\ \bibnamefont {{Aurnou}}}, \bibinfo
  {author} {\bibfnamefont {S.}~\bibnamefont {{Perrard}}}, \bibinfo {author}
  {\bibfnamefont {A.}~\bibnamefont {{Ribeiro}}}, \bibinfo {author}
  {\bibfnamefont {L.}~\bibnamefont {{Rodet}}}, \ and\ \bibinfo {author}
  {\bibfnamefont {P.}~\bibnamefont {{Le Gal}}},\ }\href@noop {} {\enquote
  {\bibinfo {title} {{Experimental study of internal wave generation by
  convection in water}},}\ } (\bibinfo {year} {2015}),\ \bibinfo {note}
  {{Submitted to {\it Fluid Dyn. Res.}}}\BibitemShut {Stop}%
\bibitem [{\citenamefont {Ascher}\ \emph {et~al.}(1997)\citenamefont {Ascher},
  \citenamefont {Ruuth},\ and\ \citenamefont {Spiteri}}]{ascher97}%
  \BibitemOpen
  \bibfield  {author} {\bibinfo {author} {\bibfnamefont {U.~M.}\ \bibnamefont
  {Ascher}}, \bibinfo {author} {\bibfnamefont {S.~J.}\ \bibnamefont {Ruuth}}, \
  and\ \bibinfo {author} {\bibfnamefont {R.~J.}\ \bibnamefont {Spiteri}},\
  }\href {\doibase 10.1016/S0168-9274(97)00056-1} {\bibfield  {journal}
  {\bibinfo  {journal} {Appl. Numer. Math.}\ }\textbf {\bibinfo {volume}
  {25}},\ \bibinfo {pages} {151} (\bibinfo {year} {1997})}\BibitemShut
  {NoStop}%
\bibitem [{\citenamefont {{Lecoanet}}\ and\ \citenamefont
  {{Quataert}}(2013)}]{lq13}%
  \BibitemOpen
  \bibfield  {author} {\bibinfo {author} {\bibfnamefont {D.}~\bibnamefont
  {{Lecoanet}}}\ and\ \bibinfo {author} {\bibfnamefont {E.}~\bibnamefont
  {{Quataert}}},\ }\href {\doibase 10.1093/mnras/stt055} {\bibfield  {journal}
  {\bibinfo  {journal} {\mnras}\ }\textbf {\bibinfo {volume} {430}},\ \bibinfo
  {pages} {2363} (\bibinfo {year} {2013})},\ \Eprint
  {http://arxiv.org/abs/1210.4547} {arXiv:1210.4547 [astro-ph.SR]} \BibitemShut
  {NoStop}%
\bibitem [{\citenamefont {{Rogers}}\ and\ \citenamefont
  {{MacGregor}}(2011)}]{rm11}%
  \BibitemOpen
  \bibfield  {author} {\bibinfo {author} {\bibfnamefont {T.~M.}\ \bibnamefont
  {{Rogers}}}\ and\ \bibinfo {author} {\bibfnamefont {K.~B.}\ \bibnamefont
  {{MacGregor}}},\ }\href {\doibase 10.1111/j.1365-2966.2010.17493.x}
  {\bibfield  {journal} {\bibinfo  {journal} {\mnras}\ }\textbf {\bibinfo
  {volume} {410}},\ \bibinfo {pages} {946} (\bibinfo {year} {2011})},\ \Eprint
  {http://arxiv.org/abs/1009.5933} {arXiv:1009.5933 [astro-ph.SR]} \BibitemShut
  {NoStop}%
\bibitem [{\citenamefont {{Zahn}}\ \emph {et~al.}(1997)\citenamefont {{Zahn}},
  \citenamefont {{Talon}},\ and\ \citenamefont {{Matias}}}]{ztm97}%
  \BibitemOpen
  \bibfield  {author} {\bibinfo {author} {\bibfnamefont {J.-P.}\ \bibnamefont
  {{Zahn}}}, \bibinfo {author} {\bibfnamefont {S.}~\bibnamefont {{Talon}}}, \
  and\ \bibinfo {author} {\bibfnamefont {J.}~\bibnamefont {{Matias}}},\
  }\href@noop {} {\bibfield  {journal} {\bibinfo  {journal} {\aap}\ }\textbf
  {\bibinfo {volume} {322}},\ \bibinfo {pages} {320} (\bibinfo {year}
  {1997})},\ \Eprint {http://arxiv.org/abs/astro-ph/9611189} {astro-ph/9611189}
  \BibitemShut {NoStop}%
\bibitem [{\citenamefont {{Press}}(1981)}]{press81}%
  \BibitemOpen
  \bibfield  {author} {\bibinfo {author} {\bibfnamefont {W.~H.}\ \bibnamefont
  {{Press}}},\ }\href {\doibase 10.1086/158809} {\bibfield  {journal} {\bibinfo
   {journal} {\apj}\ }\textbf {\bibinfo {volume} {245}},\ \bibinfo {pages}
  {286} (\bibinfo {year} {1981})}\BibitemShut {NoStop}%
\bibitem [{\citenamefont {{Taylor}}\ and\ \citenamefont
  {{Sarkar}}(2007)}]{ts07}%
  \BibitemOpen
  \bibfield  {author} {\bibinfo {author} {\bibfnamefont {J.~R.}\ \bibnamefont
  {{Taylor}}}\ and\ \bibinfo {author} {\bibfnamefont {S.}~\bibnamefont
  {{Sarkar}}},\ }\href {\doibase 10.1017/S0022112007008087} {\bibfield
  {journal} {\bibinfo  {journal} {Journal of Fluid Mechanics}\ }\textbf
  {\bibinfo {volume} {590}},\ \bibinfo {pages} {331} (\bibinfo {year}
  {2007})}\BibitemShut {NoStop}%
\bibitem [{\citenamefont {{Freund}}(2001)}]{freund01}%
  \BibitemOpen
  \bibfield  {author} {\bibinfo {author} {\bibfnamefont {J.~B.}\ \bibnamefont
  {{Freund}}},\ }\href@noop {} {\bibfield  {journal} {\bibinfo  {journal}
  {Journal of Fluid Mechanics}\ }\textbf {\bibinfo {volume} {438}},\ \bibinfo
  {pages} {277} (\bibinfo {year} {2001})}\BibitemShut {NoStop}%
\bibitem [{\citenamefont {{Boersma}}(2005)}]{boersma05}%
  \BibitemOpen
  \bibfield  {author} {\bibinfo {author} {\bibfnamefont {B.~J.}\ \bibnamefont
  {{Boersma}}},\ }\href {\doibase 10.1007/s00162-004-0107-7} {\bibfield
  {journal} {\bibinfo  {journal} {Theoretical and Computational Fluid
  Dynamics}\ }\textbf {\bibinfo {volume} {19}},\ \bibinfo {pages} {161}
  (\bibinfo {year} {2005})}\BibitemShut {NoStop}%
\bibitem [{\citenamefont {{Michaelian}}(2002)}]{michaelian02}%
  \BibitemOpen
  \bibfield  {author} {\bibinfo {author} {\bibfnamefont {M.}~\bibnamefont
  {{Michaelian}}},\ }\href {\doibase 10.1016/S0997-7546(01)01158-X} {\bibfield
  {journal} {\bibinfo  {journal} {European Journal of Mechanics B Fluids}\
  }\textbf {\bibinfo {volume} {21}},\ \bibinfo {pages} {1} (\bibinfo {year}
  {2002})}\BibitemShut {NoStop}%
\bibitem [{\citenamefont {{Brummell}}\ \emph {et~al.}(2002)\citenamefont
  {{Brummell}}, \citenamefont {{Clune}},\ and\ \citenamefont
  {{Toomre}}}]{bct02}%
  \BibitemOpen
  \bibfield  {author} {\bibinfo {author} {\bibfnamefont {N.~H.}\ \bibnamefont
  {{Brummell}}}, \bibinfo {author} {\bibfnamefont {T.~L.}\ \bibnamefont
  {{Clune}}}, \ and\ \bibinfo {author} {\bibfnamefont {J.}~\bibnamefont
  {{Toomre}}},\ }\href {\doibase 10.1086/339626} {\bibfield  {journal}
  {\bibinfo  {journal} {\apj}\ }\textbf {\bibinfo {volume} {570}},\ \bibinfo
  {pages} {825} (\bibinfo {year} {2002})}\BibitemShut {NoStop}%
\end{thebibliography}%

\end{document}